\documentclass[journal]{vgtc}                

\ifpdf
  \pdfoutput=1\relax                   
  \usepackage{graphicx}                
  \DeclareGraphicsExtensions{.pdf,.png,.jpg,.jpeg} 
\else
  \usepackage{graphicx}                
  \DeclareGraphicsExtensions{.eps}     
\fi%

\graphicspath{{figures/}{pictures/}{images/}{./}} 

\usepackage{microtype}                 
\PassOptionsToPackage{warn}{textcomp}  
\usepackage{textcomp}                  
\usepackage{mathptmx}                  
\usepackage{times}                     
\usepackage{cite}                      
\usepackage{tabu}                      
\usepackage{booktabs}                  

\usepackage{algorithm}
\usepackage{algorithmicx}
\usepackage{algpseudocode}
\usepackage{amsmath}
\usepackage{enumitem}
\usepackage{euscript}
\usepackage{wrapfig}

\newcommand{\myparagraph}[1]{\mbox{\ } \newline \noindent \textbf{#1}}
\renewcommand{\paragraph}[1]{\myparagraph{#1}}

\newif\ifshowcomments
\showcommentstrue

\ifshowcomments
\newcommand{\TODO}[1]{{\color{red}{[TODO: #1]}}}
\newcommand{\yh}[1]{{\color{black}{#1}}}
\newcommand{\phil}[1]{{\color[rgb]{0.9,0.1,0.1}{[PF: #1]}}}
\newcommand{\od}[1]{{\color{red}#1}}
\newcommand{\dl}[1]{{\color{magenta} [DL: #1]}}
\newcommand{\mf}[1]{{\color{blue} [MF: #1]}}

\else
\newcommand{\TODO}[1]{}
\newcommand{\revised}[1]{}
\newcommand{\yh}[1]{}
\newcommand{\phil}[1]{}
\newcommand{\od}[1]{}
\newcommand{\dl}[1]{}
\newcommand{\mf}[1]{}
\fi

\onlineid{1326}

\vgtccategory{Research}
\vgtcpapertype{algorithm/technique}

\usepackage{color}

\title{Palettailor: Discriminable Colorization for Categorical Data
}

\author{Kecheng Lu, Mi Feng, Xin Chen, Michael Sedlmair, Oliver Deussen, Dani Lischinski, Zhanglin Cheng, Yunhai Wang }
\authorfooter{
\item
K. Lu, X. Chen, Y. Wang are with Shandong University. E-mail: \{lukecheng0407, chenxin199634, cloudseawang\}@gmail.com.
\item
M. Feng is with Twitter Inc. E-mail: miamfeng@gmail.com.
\item
K. Lu and Z. Cheng are with Shenzhen VisuCA Key Lab, SIAT, China. E-mail: zl.cheng@siat.ac.cn.
\item
M. Sedlmair is with VISUS, University of Stuttgart, Germany. E-mail: michael.sedlmair@visus.uni-stuttgart.de.
\item
O. Deussen is with Konstanz University, Germany and Shenzhen VisuCA Key Lab, SIAT, China. E-mail: oliver.deussen@uni-konstanz.de.
\item
D. Lischinski is with Hebrew University, Jerusalem, Israel. E-mail: danix@mail.huji.ac.il
\item
Y. Wang and Z. Cheng are joint corresponding authors.
}

\shortauthortitle{Biv \MakeLowercase{\textit{et al.}}: Palettailor: Discriminable Colorization for Categorical Data}

\abstract{%
We present an integrated approach for creating and assigning color palettes to different visualizations such as multi-class scatterplots, line, and bar charts.  While other methods separate the creation of colors from their assignment, our approach takes data characteristics  into account to produce color palettes, which are then assigned in a way that fosters better visual discrimination of classes. To do so, we use a customized optimization based on simulated annealing  to maximize the combination of three carefully designed color scoring functions: point distinctness, name difference, and color discrimination. We
compare our approach to state-of-the-art palettes with a controlled user study for scatterplots and line charts, furthermore we performed a case study.
Our results show that Palettailor, as a fully-automated approach, generates color palettes with a higher discrimination quality than existing approaches.
The efficiency of our optimization allows us also to incorporate user modifications into the color selection process.
} 


\CCScatlist{ 
 \CCScat{K.6.1}{Management of Computing and Information Systems}%
{Project and People Management}{Life Cycle};
 \CCScat{K.7.m}{The Computing Profession}{Miscellaneous}{Ethics}
}

\teaser{
  \centering
  \includegraphics[width=\linewidth]{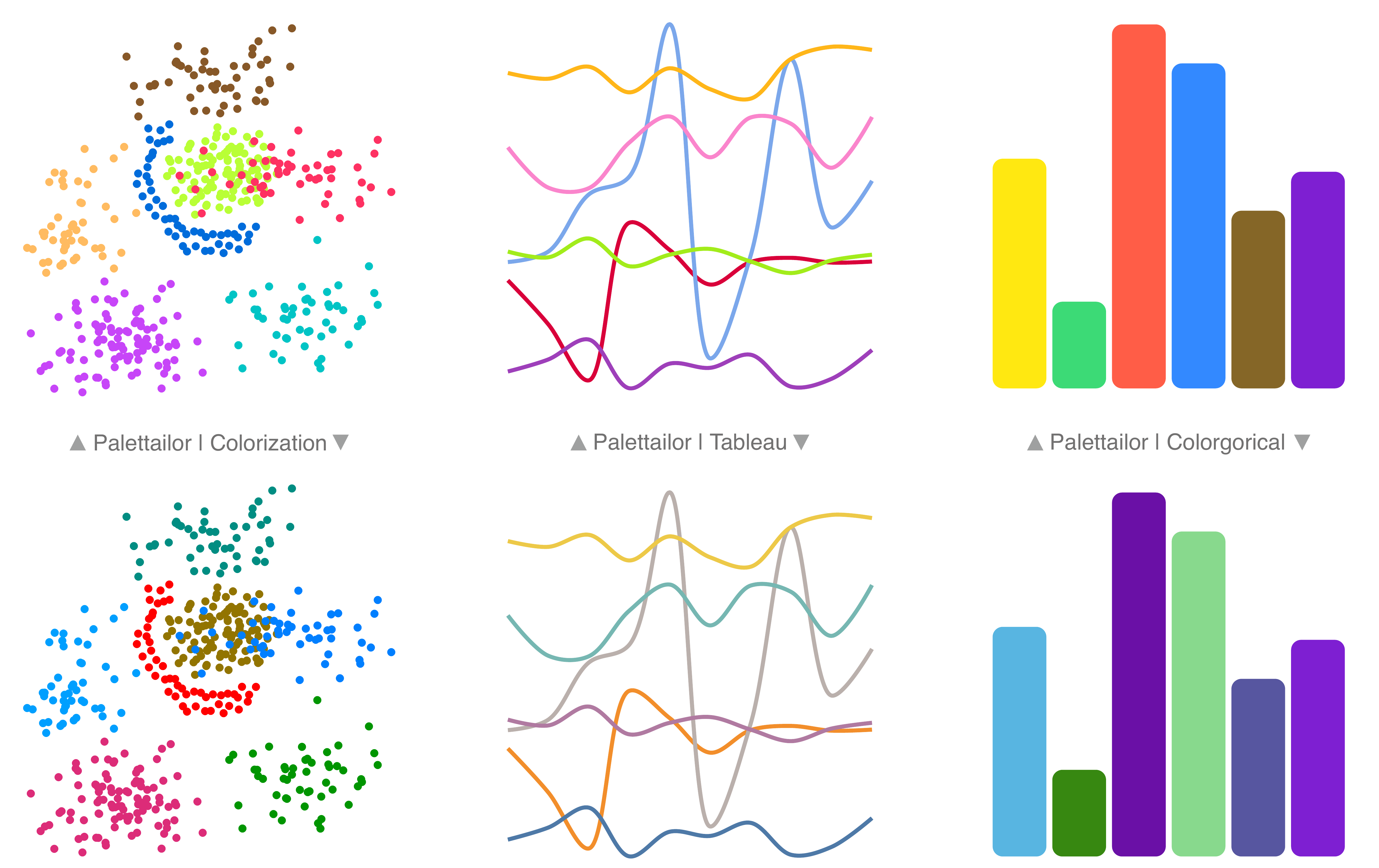}
  \caption{
  Results for different types of categorical data visualizations: (left) Palettailor versus Colorization\cite{chen2014visual}; (center) Palettailor versus Tableau\cite{tableau}; (right) Palettailor versus Colorgorical\cite{gramazio2017colorgorical}. Our system integrates the creation and the assignment of colours to a visualization in a data-aware manner.
  }
	\label{fig:teaser}
}

\vgtcinsertpkg

\begin{document}


\maketitle

\section{Introduction}

Visualizing categorical data in statistical graphics such as  bar charts, line charts, or scatterplots is most commonly realized by encoding each category (or class) with a unique color. One major task during visual analysis is then to discriminate between the different  classes. While it is well-known that  class discriminability is strongly influenced by the assigned colors~\cite{gleicher2013perception,lee2013perceptually}, finding an appropriate set of colors for the different classes in a specific visualization is still a complex and time-consuming endeavor, even for experts.

The most common way to obtain an appropriate color mapping is to find a good color palette first and then assign the colors to classes in the best possible way.
To ease this procedure, a few color palette tools have been provided, such as ColorBrewer~\cite{harrower2003colorbrewer} or Colorgorical~\cite{gramazio2017colorgorical}, which allow users to select highly discriminable and preferable palettes. Since the creation of such palettes ignores the specific data of a visualization, a good palette might still not be optimal to visually discriminate classes in different forms of visualization. Hence, users often need to try different palettes and color assignment schemes until the desired result is achieved. Recently, Wang et al.~\cite{wang2019optimizing} proposed a method that automatically assigns colors of a given palette to classes of multi-class scatterplots by maximizing their discriminability.  This technique enables users to bypass the second stage of the standard color assignment process, but it is limited to scatterplots and still requires the author to select a good palette. In contrast, Chen et al.~\cite{chen2014visual} proposed an automatic color
selection approach \yh{for multi-class} scatterplots by searching discriminative colors in the a* and b* channel of the CIELAB space. However, 
leaving out L* channel often does not allow to find colors with high enough discriminability, especially when the number of classes is large. \yh{Since such an approach directly colorizes multi-class scatterpots without any input palette, we refer to it as \textit{Colorization} in this paper.}

To fill this gap, we propose Palettailor, a data-aware color palette generation framework, that automatically generates categorical palettes with maximized discriminability for different visualization types. To achieve this goal, we adapted three color-scoring functions: Point Distinctness~\cite{wang2019optimizing}, Name Difference~\cite{heer2012color},  and the Color Discrimination constraint. The first function is measured from the visualizations, while the last two terms from the palettes. By doing so, we combine the two stages of palette selection and color assignment to a single framework. Users only need to specify weights of the different color-scoring functions and background color to create suitable palettes. \yh{In terms of these functions, our notion of ``data-aware'' mainly refers to the spatial distribution and cardinality of categories in a given scatterplot, while data semantics can be reflected by, e.g., constraining color names (see Section~\ref{sec:constraint}). As far as we know, few existing color palette generation tools take all these factors into account.}

To realize this idea, we customize state-of-the-art class separation measures~\cite{aupetit2016sepme} for designing color-based class discriminability. Wang et al.~\cite{wang2019optimizing} measure Point Distinctness by using a K-Nearest-Neighbor-Graph with taking the two nearest neighbors per point, but we hypothesize that this might not work for all cases since the KNN method would only connect with $k$ neighbors and might not be able to accurately characterize local class separability.
Instead, we suggest to use an
$\alpha$-shape~\cite{veltkamp1992gamma} graph that connects a set of points within an $\alpha$-ball so as to better characterize the local discriminability. 

With such a pre-computed geometry-based class separation, an optimal color palette can be found by evaluating all possible solutions and then ranking them accordingly. The color space, however, is too large for an exhaustive investigation and so we propose a customized simulated annealing algorithm~\cite{pham2012intelligent} to rapidly and efficiently obtain a near-optimal solution. In doing so, we can generate the palettes for scatterplots and other types of visualizations with 40 classes in less than 15 seconds.

We evaluated our approach through carefully designed scatterplots and line charts by comparing our colorized results with the ones produced by the state-of-the-art palettes (e.g.,Tableau~\cite{tableau} and Colorgorical).
We first conducted a pilot study for each visualization type to justify whether our design is valid.
We then carried out an online study with three tasks for a scatterplot: the first two tasks investigated how well our generated palettes help users to discriminate classes, while the third task aimed at finding out if our results align with user preferences. Similarly, we carried out another online study for a line chart with a discrimination task and again measuring the alignment with user preferences. The results show that our method is able to produce palettes optimized for class discrimination and aligned with user preferences for different chart types that are \yh{more effective than state-of-the-art palettes in most cases}.

On the basis of our method, we implemented a web-based palette generation tool that allows users to generate suitable palettes for multi-class scatterplots, but also for other categorical visualizations such as bar and line charts. In summary, the main contributions of this paper are:
\begin{itemize}
\vspace*{-2mm}
\item
We propose a data-aware approach based on simulated annealing for automatically generating color palettes for different visualization types.  It is based on user-defined weights for three different scoring functions
and extends state-of-the-art separation measures (Section~\ref{sec:method}).

\vspace*{-2mm}
\item
We quantitatively evaluate the resulting color palettes and visualizations, and compare the results with state-of-the-art color palettes; online user studies show the usefulness of our approach (Section~\ref{sec:evaluation}).

\vspace*{-2mm}
\item
We present an interactive tool\footnote{\small \url{http://www.palettailor.net}} that demonstrates the practical applicability of our method and  helps exploring  categorical information visualizations (Section~\ref{sec:interaction}).
\end{itemize}

\section{Related Work}
Since we focus on color palette generation, related work is divided into color palette creation and color palette optimization. \yh{For the generation of color ramps, a slightly different color assignment problem, we refer readers to recent
surveys~\cite{tominski2008task,wijffelaars2008generating,Mittelstadt2015Color-32461,nardini2019making} .}


\subsection{Color Palette Creation}
\label{sec:colordesign}
Creating a categorical color palette that enables viewers to distinguish between data elements is a crucial task for many visualizations. Previous attempts have derived practical guidelines based on perceptual constraints, such as ``colors should be well separated'' \cite{healey1996choosing}, ``should not compete with each other''\cite{tufte1990envisioning}, and ``should not be unappealing''\cite{zeileis2009escaping}. Following these guidelines, a few interactive color palette creation systems have been developed. Healey~\cite{healey1996choosing} proposed to choose representative colors from ten hue regions in Munsell's color space, while maximizing the perceptual distances between them. Maxwell~\cite{maxwell2000visualizing} selects colors that maximize the discriminability between classes, but also takes into account their spatial distribution. \yh{Their approach has two major limitations. One is that the generated colors are not aesthetically pleasing or easy to see; and the other is that it defines perceptual distances based on the so-called maximum scaled difference, which might not align well with human perception.}
We want to find a much more general solution. Colorgorical~\cite{gramazio2017colorgorical} also overcomes this drawback, but does not take into account the data distribution. The resulting palettes might thus not allow to discriminate classes in given data or in a particular visualization. \yh{The Colorization method proposed by Chen et al.~\cite{chen2014visual} automatically selects colors for multi-class scatterplots} in the  CIELAB color space to yield a maximal color distinguishability. This technique targets the same problem as ours, but has three main drawbacks: (i) using density to encode class separability does not lead to an accurate measurement; (ii) leaving L* as a user-adjustable parameter and only optimizing a* and b* values limits the approach to a 2D plane in CIELAB and does not provide enough colors with large color differences; (iii) using Euclidean Distance for computing color distances leads to generated colors that are often at the boundary of the 2d plane due to the optimization. Our method avoids these drawbacks.

Another approach for designing discriminable palettes is to use prefabricated ones. A typical example is ColorBrewer~\cite{harrower2003colorbrewer}, an online tool for generating color palettes optimized for choropleth maps. Although ColorBrewer offers different high-quality palettes for different cardinalities, it does not allow users to customize them. Colorgorical~\cite{gramazio2017colorgorical} enables users to customize palettes by specifying desired hues, but it does not take the underlying data into account and thus cannot customize colors for specific classes or visualizations. Our data-aware approach inherently offers such flexibility.

\subsection{Color Palette Optimization}
Once a color palette is selected, users often want to refine the colors until a desired set is achieved and then find a proper assignment to the categories within the data.

\vspace*{-5pt}
\paragraph{Color Optimization.} \
Optimizing a selected color palette can be performed by applying different criteria, such as optimizing data comprehension, aesthetics, energy saving, or helping with color vision deficiency (CVD)~\cite{zhou2016survey}. Wang et al.~\cite{wang2008color} introduced a knowledge-based system that employs established color design rules to optimize chosen colors for discriminating spatial structures in 2D and 3D visualizations. Chuang et al.~\cite{chuang2009energy} proposed to optimize color palettes with the goal of lowering the energy consumption of display devices, and Machado et al.~\cite{machado2009physiologically} presented a physiologically-based model for simulating color perception so that the refined palettes can efficiently improve the visualization experience for individuals with CVD.

Lee et al.~\cite{lee2013perceptually} proposed to measure the visual saliency of each visible point of a class and used this metric to optimize palettes for better class discrimination.
This method takes the color contrast against the background into account, however, \yh{it aims to optimize the given color palette in terms of  spatial adjacency of the given data.} In contrast to their work, our approach directly produces discriminable colors for different categorical visualization types such as bar or line charts; furthermore, we also incorporate color contrast with the background into the discriminability measure.

Rather than optimizing palettes for a specific visualization, Fang et al.~\cite{fang2017categorical} provided an approach for maximizing perceptual distances among a set of given colors. Although this approach can incorporate various user-defined constraints, it is independent of the given data and thus the produced visualization might not clearly show the different class structures in the data.

\vspace*{-5pt}
\paragraph{Color Assignment.} \
Assigning colors properly to different classes is a crucial step in categorical data visualization. Assuming that categorical values have semantics, Lin et al.~\cite{lin2013selecting} propose a method that automatically matches each value to a unique color from a given palette. \yh{Setlur and Stone~\cite{setlur2016linguistic} use linguistic information to generate semantically meaningful colors.}  Since most classes in a visualization might not have clear semantics, Wang et al.~\cite{wang2019optimizing} allow user modification and customization within their optimization framework. \yh{To that end, they take spatial adjacency into account, but require a discriminable palette first.} Our approach unifies palette creation and color assignment into a single procedure and is also able to adapt to user wishes. In addition, it supports palette completion and introduces semantics by handling color names. \yh{By constraining color selection with user-specified color names, our approach could also be extended to achieve data semantics-based colorization similar to existing approaches~\cite{lin2013selecting,setlur2016linguistic}.}

\section{Methods}
\label{sec:method}
Given categorical data with $m$ classes $C=\{C_1,\cdots,C_m\}$, we assume to have a point set  $\mathbf{X}=\{\mathbf{x}_1,\cdots , \mathbf{x}_n\}$ where each $\mathbf{x}_i$ is associated with a class label $l(\mathbf{x}_i)$ and the $j$-th class (with $n_j$ data points) consist of $\{\mathbf{x}_1^j, \cdots , \mathbf{x}_{n_j}^j\}, j \in M = \{ 1, \cdots, m \} $,  the background color is represented by $c_0$. We formulate the search for a color palette $P=\{c_1,\cdots,c_m\}$ as an optimization problem with the \textbf{objective function} $E(P)$:
\begin{align} 
	\mathop{\arg\max}_{P}E(P)= \omega_0 E_{PD}+\omega_1 E_{ND}+\omega_2 E_{CD},
\label{eq:objectivefunction}
\end{align}
where each class $C_i$ is assigned a unique color $c_i$ and each weight $\omega_i$ is a value range from 0 to 1.

In the following, we will introduce the constraints we use for optimizing the colors in different visualization types: Point Distinctness(PD), Name Difference(ND) and Color Discrimination(CD), and then describe how we solve the overall optimization problem.

\begin{figure}[h]
	\centering
	\includegraphics[width=0.8\linewidth]{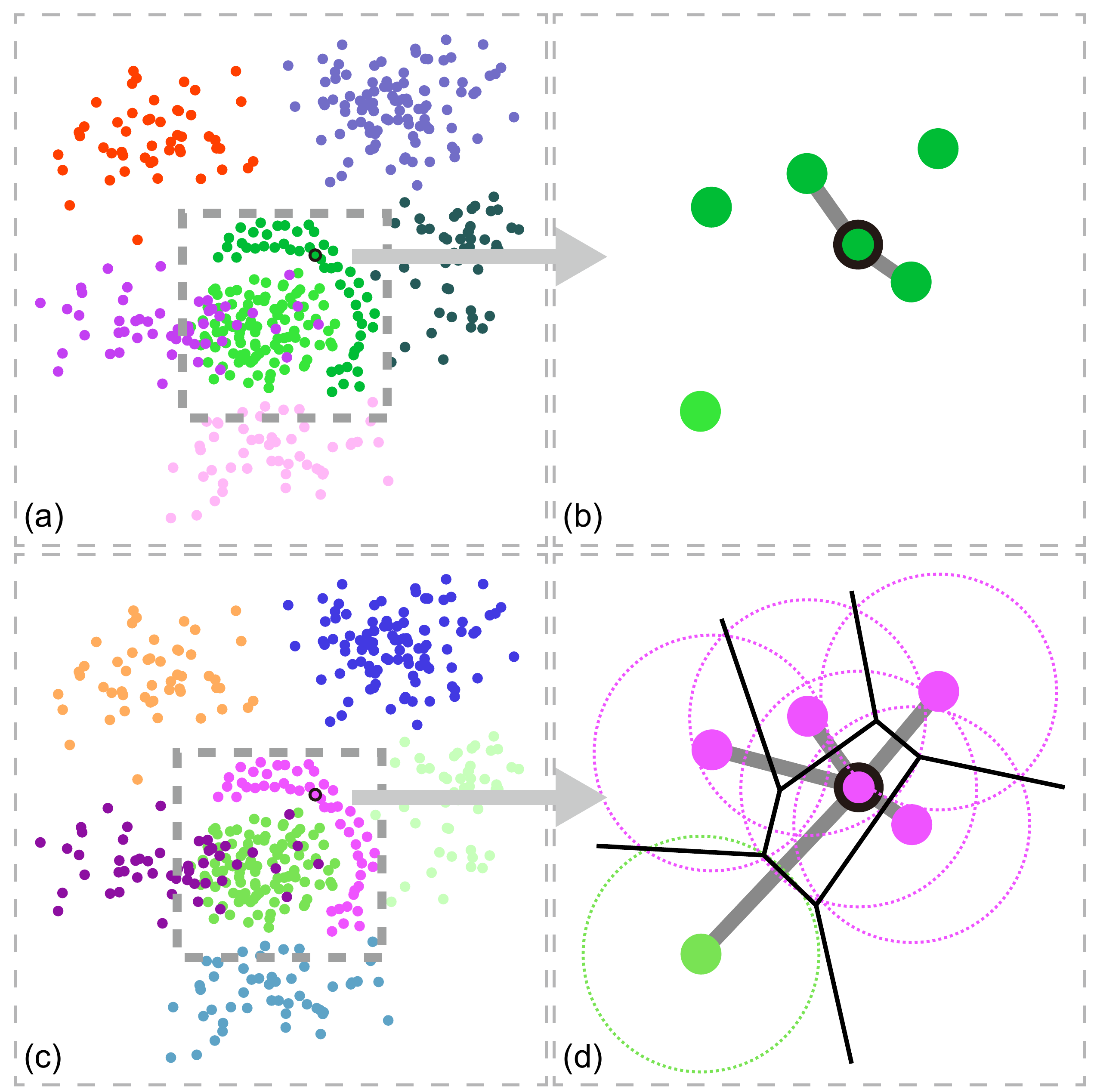}
	\caption{Comparison of nearest neighbors definitions  in KNN graphs and $\alpha$-Shape graphs. (a) Results generated by a KNN graph using only point distinctness: the generated colors are hard to discriminate; (b) Nearest neighbors of the selected point in the KNN graph; (c) Results generated by $\alpha$-Shape graph only using point distinctness: the generated colors are easily to discriminate;
		(d) Nearest neighbors of the selected point in the $\alpha$-Shape graph.}
	\label{fig:alphashape}
	\vspace{-4mm}
\end{figure}

\subsection{Scoring Functions}
\label{sec:factors}

Since scatterplots are the most commonly used chart type with points as visual marks, we decided to use them to illustrate our method. The arguments, however, also hold for other types of visualizations, as shown later.

\vspace{-5pt}
\paragraph{Point Distinctness.}\
Similar to \cite{wang2019optimizing}, we define the distinctness of a data point $x_i$ from its neighbours using a color mapping $P$ as follows:
\begin{align}\label{eq:pd}
 \alpha (\mathbf{x}_i)=\frac{1}{|\Omega_i|} \sum_{\mathbf{x}_j \in \Omega_i} \Delta \epsilon(c_i,c_j)g(d(\mathbf{x}_i,\mathbf{x}_j)) ,
\end{align}
where the $\Omega_i$ are the nearest neighbors of $\mathbf{x}_i$ with $c_i=P(l(\mathbf{x}_i)), c_j=P(l(\mathbf{x}_j))$ color mapping functions, $\Delta \epsilon$ is the perceptual color distance metric called CIEDE2000~\cite{sharma2005ciede2000}  and $g(d(\mathbf{x}_i,\mathbf{x}_j))$ is a distance-based function that assigns large weights to nearby points and small weights to distant points; we used $g(d)=\frac{1}{d}$ in our implementation.

In contrast to Wang et al.~\cite{wang2019optimizing}, we define $\Omega_i$ using an $\alpha$-Shape graph~\cite{aupetit2016sepme} instead of a KNN graph. The reason for this is that nearest neighbors within of an KNN graph may have an arbitrarily large or small distance to the point of interest. Although the distance-based function $g(d)$ helps for reducing the influence of  neighbors with large distances, having neighbors with only small distances might result in a colorization that is hard to discriminate. Fig.~\ref{fig:alphashape}(a) shows an example where the two well-separated clusters enclosed by the rectangle receive similar colors for the KNN-based approach and are thus hard to distinguish. In contrast to this, the $\alpha$-Shape graph connects only points that are neighbors in the Delaunay graph and in addition their $\alpha$-balls intersect (see Fig.~\ref{fig:alphashape}(d)). This property allows us to select the closest neighbors of a given point within a certain radius. Fig.~\ref{fig:alphashape}(c) shows the colorization result, now the two nearby clusters enclosed by the rectangle can be discriminated easily.

The colorization results for the same scatterplot using these two different graphs are shown in~\autoref{fig:alphashape}(a,c).Since the two nearby clusters highlighted by the rectangle are well separated and dense, the $k$ nearest neighbors of each point belong to the same class in the KNN graph  and the point distinctness is zero; this leads to similar assigned colors for the two clusters. In contrast, point distinctness computed from the $\alpha$-Shape graph takes into account only neighbors within a given radius and therefore results a discriminable colorization.
%

Point Distinctness is defined as the sum of the Point Distinctness of all points in each class:

\begin{align}\label{eq:Epd}
 E_{PD} = \sum_{j=1}^{m} \sum_{i=1}^{n_j} \alpha(\mathbf{x}_i^j).
\end{align}

\begin{figure}[tbh]
	\centering
	\includegraphics[width=0.8\linewidth]{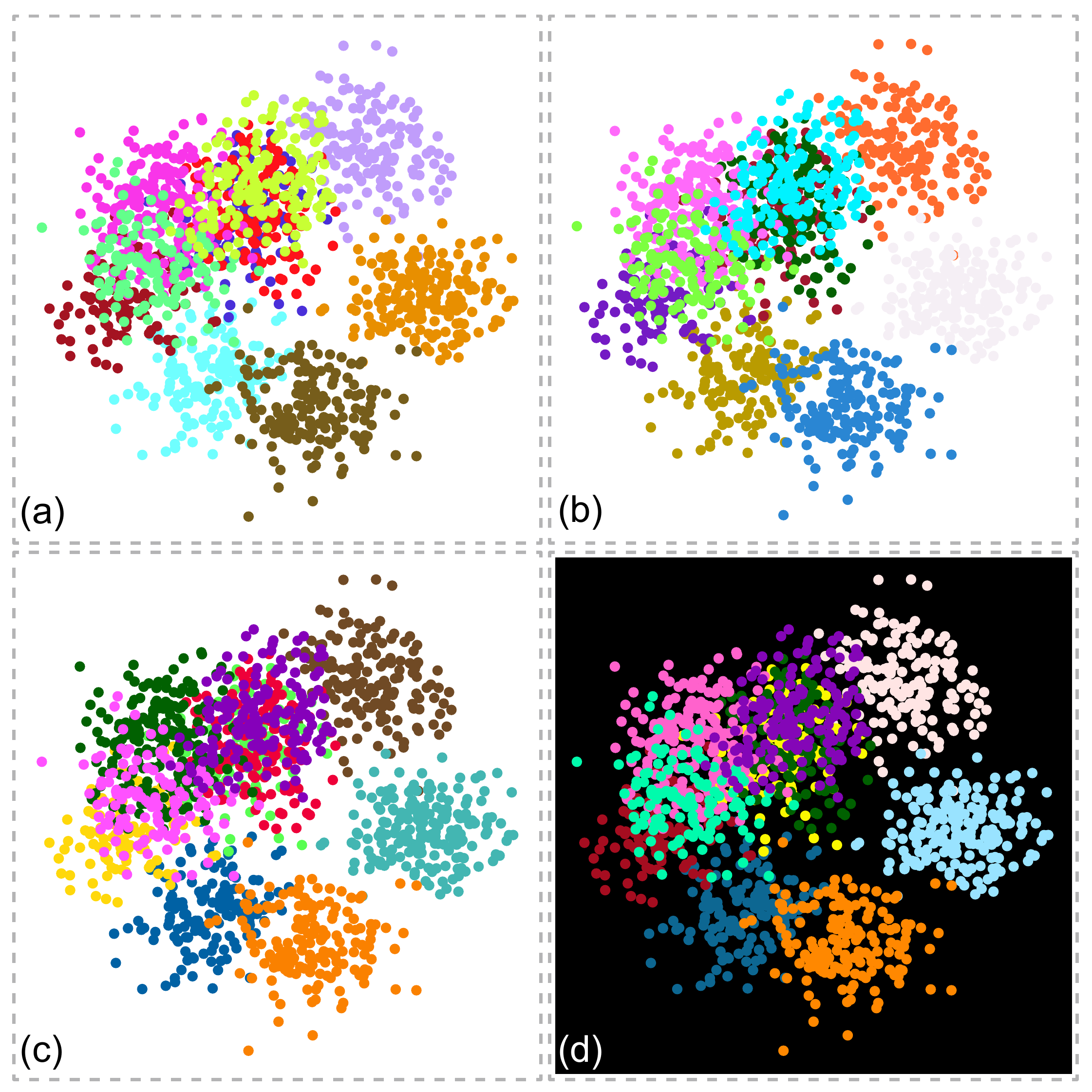}
	\caption{Colorization results based on different optimization criteria:
		(a) without Class Discrimination constraint;
		(b) without considering background color making it hard to see points of one class;
		(c)  integrating the white background  into the optimization lets the class appear in a dark color;
		(d)  integrating the black background color  into the optimization changes the colorization of many classes in order to create the necessary contrast.}
	\label{fig:discriminationConstraint}
	\vspace{-4mm}
\end{figure}

\paragraph{Name Difference.}\
Two colors might be perceptually different but are referenced by the same name, confusing color selection. For example,
the color ``Violet'' is perceptually different from ``Dark Purple'', but they are often referred by the same name ``Purple'',
making it difficult to reference them by name in a visualization. To model such name-color associations, Heer and Stone\cite{heer2012color} introduce the concept of Name Difference (ND), that measures the probability of two colors having the same name. It was used by Colorgorical\cite{gramazio2017colorgorical} for generating palettes with perceptually separated colors. For a palette, its ND is defined as the minimal Hellinger distance\cite{heer2012color} between the probability distributions of all color name pairs. Another option is to compute the cosine distance, the angle between the two distributions:
\begin{align}\label{eq:nd}
 ND(c_i,c_j) = 1-cos(T_{c_i},T_{c_j})=1-\frac{T_{c_i}T_{c_j}}{||T_{c_i}||||T_{c_j}||}  ,
\end{align}
Where $T$ is the name count matrix for the color terms, with $T_c$ the row vector for color $c$. Due to its simplicity and familiarity, we use the cosine distance to measure the name difference of two colors.
For a palette, our goal is to compose it as much as possible from colors with different color names, thus its overall Name Difference is defined as the average value of all color pairs:
\begin{align}\label{eq:nd}
 E_{ND} = \frac{2}{m(m-1)}\sum_{i \neq j \in M}ND(c_i,c_j).
\end{align}

\paragraph{Color Discrimination.}
Since the spatial distance between marks might influence the perception of color differences \cite{brychtova2017effect}, we employ a hard constraint into our optimization: each two colors should have a minimal CIEDE2000 distance of 10, so that each pair of colors can be distinguished even for large spatial distances. Meanwhile, the minimum color distance should be as large as possible to protect the optimization from reaching unsatisfying results. Fig.~\ref{fig:discriminationConstraint}(a) shows that without using the color discrimination constraint classes with small overlap might be assigned  improper colors: the green  and  limegreen classes are hard to distinguish, also the green and cyan classes. In addition, we also take the background color $c_0$ into consideration to prevent the optimization from generating colors indistinguishable from the background, see Fig.~\ref{fig:discriminationConstraint}(b),(c),(d) for illustration. The formal definition of the Color Discrimination is:
\begin{equation}
	E_{DC}=\min_{0\leq i\neq j \leq m} \Delta \epsilon (c_i,c_j).
\end{equation}
\

To achieve good optimization results we need to balance the three terms. For the Point Distinctness, we use a normalized weight relative to an initial score at the beginning of the optimization taken from a random palette that satisfies basic requirements.
Based on out experiments Name Difference is multiplied by a factor of 2.0, while the Color Discrimination constraint is multiplied with a factor of 0.1.

\subsection{Extension to Line and Bar Charts}
\label{sec:extension}

We extended our method to generate color palettes for other categorical visualizations such as line and bar charts. We achieved this by interpreting these charts as special cases of scatterplots, which allowed us to fit them into our optimization framework. We are well aware of the fact that this is a very simple form of implementation, but the results are promising.

\begin{figure}[htb]
	\centering
	\includegraphics[width=\linewidth]{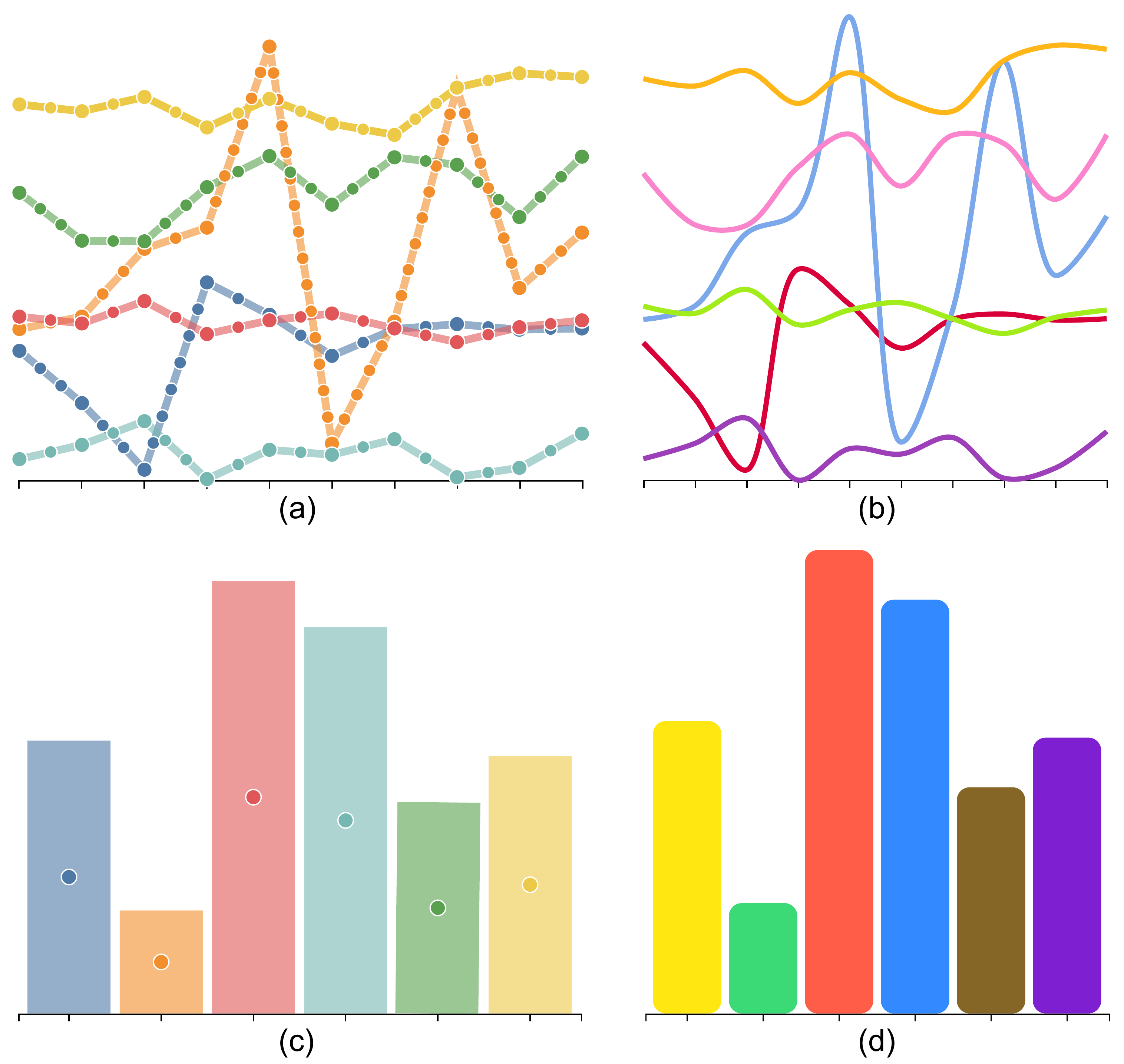}
	\vspace{-6mm}
	\caption{Converting line and bar charts to point-based representation for colorizing them: (a) each curve in a line chart is discretized into equidistant points; (b) colorized line chart;  (c) for a bar chart, the center of each bar is represented as a point and connected to its two adjacent bars for forming the graph;  (d) colorized bar chart.}
	\label{fig:extension}
	\vspace{-2mm}
\end{figure}

Fig.~\ref{fig:extension} illustrates the process. To colorize a \emph{line chart}, we discretize the lines into equidistant points and process them the same way as the points of a scatterplot.  Note that we discretize line segments with large slopes result into more point samples, indicating that such lines need to be discriminated stronger from each other. 

A \emph{bar chart} is different from a line chart in that nearest neighbors are pre-defined by the adjacent bars. Thus we treat every bar as a single point class with a center at the middle of the bar; its nearest neighbors are the two adjacent bars. An example is shown in Fig.~\ref{fig:extension}(c), where we took the length of the connecting lines between each two center points as the distance of the corresponding bars.
In addition, we used the reverse function $g(d)=\frac{1}{d}$ to scale up distances of bars with similar heights for better distinguishing them.
Our technique achieves high-quality results compared to state-of-the-art palettes and is more flexible and steerable, some results can be seen in Fig.~\ref{fig:extension}(b,d). Due to the space limits, we show more results in the supplemental material.

\begin{figure}[htb]
	\centering
	\includegraphics[width=0.9\linewidth]{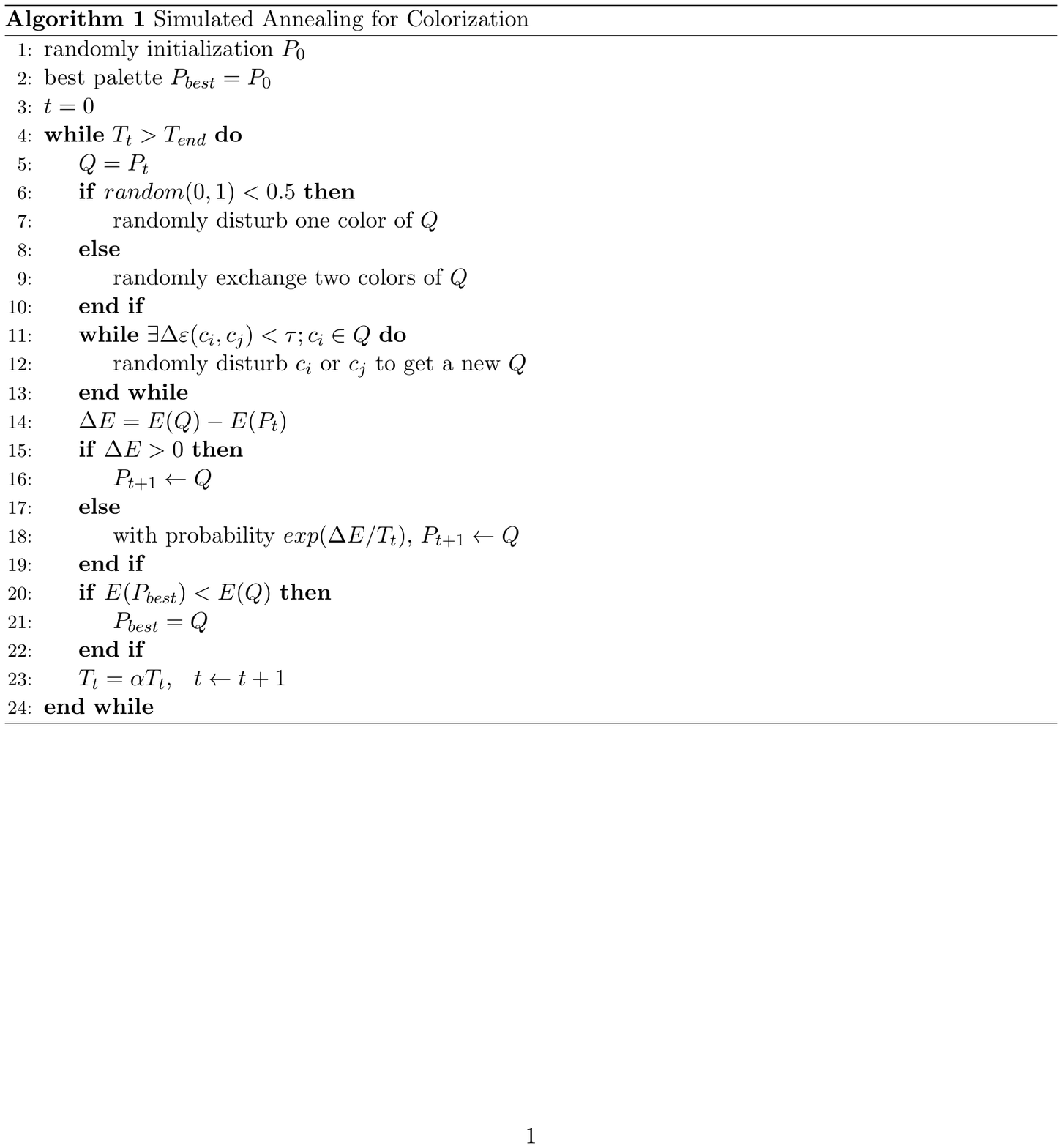}
	\label{fig:algorithm}
	\vspace{-4mm}
\end{figure}

\subsection{Simulated Annealing}
\label{sec:optimization}
Our method takes categorical data with $m$ classes as input, then iteratively tries to find an optimized color palette, with each color assigned to a class. Similar to Gramazio et al. \cite{gramazio2017colorgorical}, we filter out strongly disliked colors whose luminance value is in the range of $L\in[35,75]$ and hue values $H\in[85^\circ,114^\circ]$. To find $m$ colors with maximal energy with respect to Eq.~\ref{eq:objectivefunction}, we use Simulated  Annealing~\cite{aarts1989stochastic}, a stochastic optimization method for finding optima. In comparison to the optimization adopted by Pham and Karaboga~\cite{pham2012intelligent}, Simulated Annealing yields better results at comparable run times, this way also facilitating  an interactive generation of palettes.

Starting with a high ``temperature'' and an initial guess, the method iteratively updates the palette, while gradually lowering the temperature until convergence is achieved. During this procedure, intermediate results worse than the current iteration can be accepted with a probability related to the ``temperature''. If the cooling of this temperature is slow, it is more likely to reach the global optimum. Algorithm 1
shows the details of our implementation. It has three customized components: generating a new solution, palette refinement by imposing hard constraints, and finding the optimal colorization.
In the following we explain each of these components in detail. In line with suggestion of previous works~\cite{kirkpatrick1983optimization}, we use the following values: cooling coefficient $\alpha=0.99$, initial temperature  $T=100,000$  and end temperature  $T_{end}=0.001$. \yh{The effect of different parameter values and number of iterations on the final result is shown in the supplemental material.}

\vspace*{-5pt}
\yh{
\paragraph{Initialization of P (line 1).}\
We found that using random initialization can generate reasonable results in most cases.
Although $P$ can also be initialized by any existing good palette, the results are almost the same as the ones produced from random initializations. Since existing palettes also might not have enough colors, $P_0$ is set to a random initialization by default.}

\vspace*{-5pt}
\paragraph{Generating a new solution (line 6-10).}\
With the initial palette $P_0 =\{c_{1}$,$c_{2}$,...,$c_{m}$\}, we create the new solution by two different ways: adding a small random offset to each component of the randomly selected color $c_i$ in $P$, or by randomly exchanging two colors. These two ways of updating unify the generation of the palette and the color assignment in a single process that assigns each class its most appropriate color.

\vspace*{-5pt}
\paragraph{Palette refinement by imposing hard constraints (line 11-13).}\
To guarantee that all colors in $Q$ are sufficiently discriminable, we check if any pair of colors would be closer to each other than a noticeable difference threshold $\tau$, which is suggested to be 10, cf. ~\cite{brychtova2017effect}. If this is the case, we randomly perturb these colors until the differences between all color pairs are larger than $\tau$.

\vspace*{-5pt}
\paragraph{Finding the optimal colorization (line 14-22).}\
After palette refinement, we score the created palette $Q$ as well as the previous one $P_t$ (line 14) to decide whether to accept this solution or not (line 15-19). Our goal is to generate a discriminable visualization and hence we preserve the best result for each iteration (line 20-22).

\subsection{Time Complexity}\label{sec:param}
We implemented our algorithm using JavaScript and tested it on a computer with an Intel Core i7-7700HQ processor with 32GB memory. The point distinctness can be decomposed into two steps~\cite{wang2019optimizing}; thus, the input data can be pre-computed and we separate this pre-processing step from the simulated annealing algorithm. In this way, the performance of our method is not related to the number of data points but instead depends only on the number of classes.

\begin{figure}[thb]
	\centering
	\includegraphics[width=0.9\linewidth]{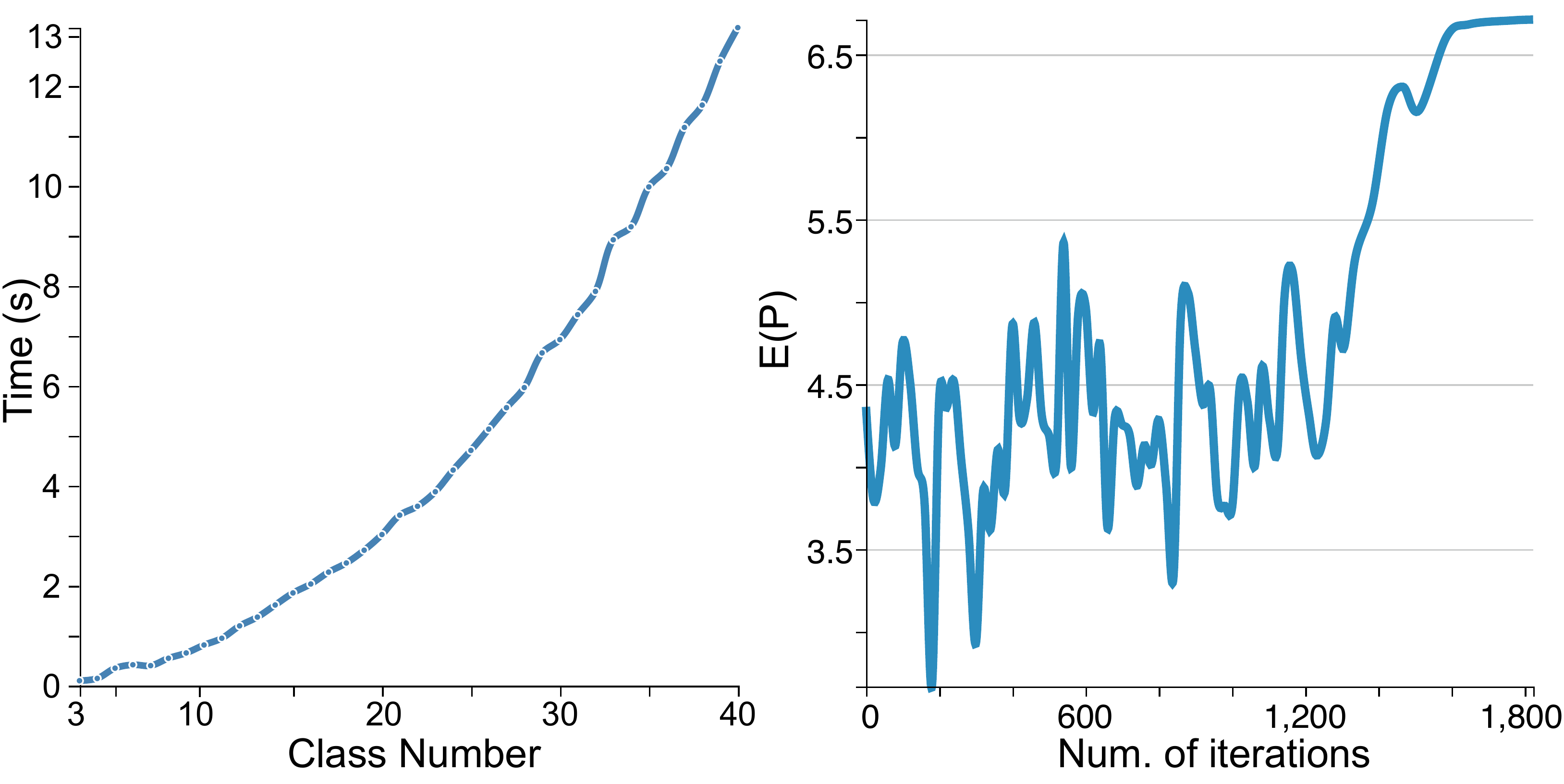}
	\caption{Time complexity of our method: (left) optimization time versus number of Classes.
		(right) energy level $E(P)$ versus number of iterations.
	}
	\label{fig:timeVSclassnum}
	\vspace{-4mm}
\end{figure}

In each iteration of the optimization we apply the scoring functions two times, with a time complexity of $O(m^2)$. For a total number of iterations $t$, the time complexity for the whole algorithm is $O(tm^2)$.
Fig.~\ref{fig:timeVSclassnum} left shows that the optimization can find reasonable palettes for scatterplots with 20 classes in less than 3s and for 40 classes in less than 15s. Fig.~\ref{fig:timeVSclassnum} right shows the convergence curves, our method has a number of strong oscillations at the beginning and then oscillates more smoothly until it converges. \yh{We checked all generated palettes with 40 classes and found that the distances of all color pairs were larger than the noticeable
difference threshold~\cite{brychtova2017effect}. Thus, we assume that Palettailor can generate 40 or more discriminable colors, of course only under the assumption of side-by-side comparison~\cite{Munzner:2001992}.}

\section{Evaluation}
\label{sec:evaluation}

\begin{figure}[htb]
	\centering
	\includegraphics[width=\linewidth]{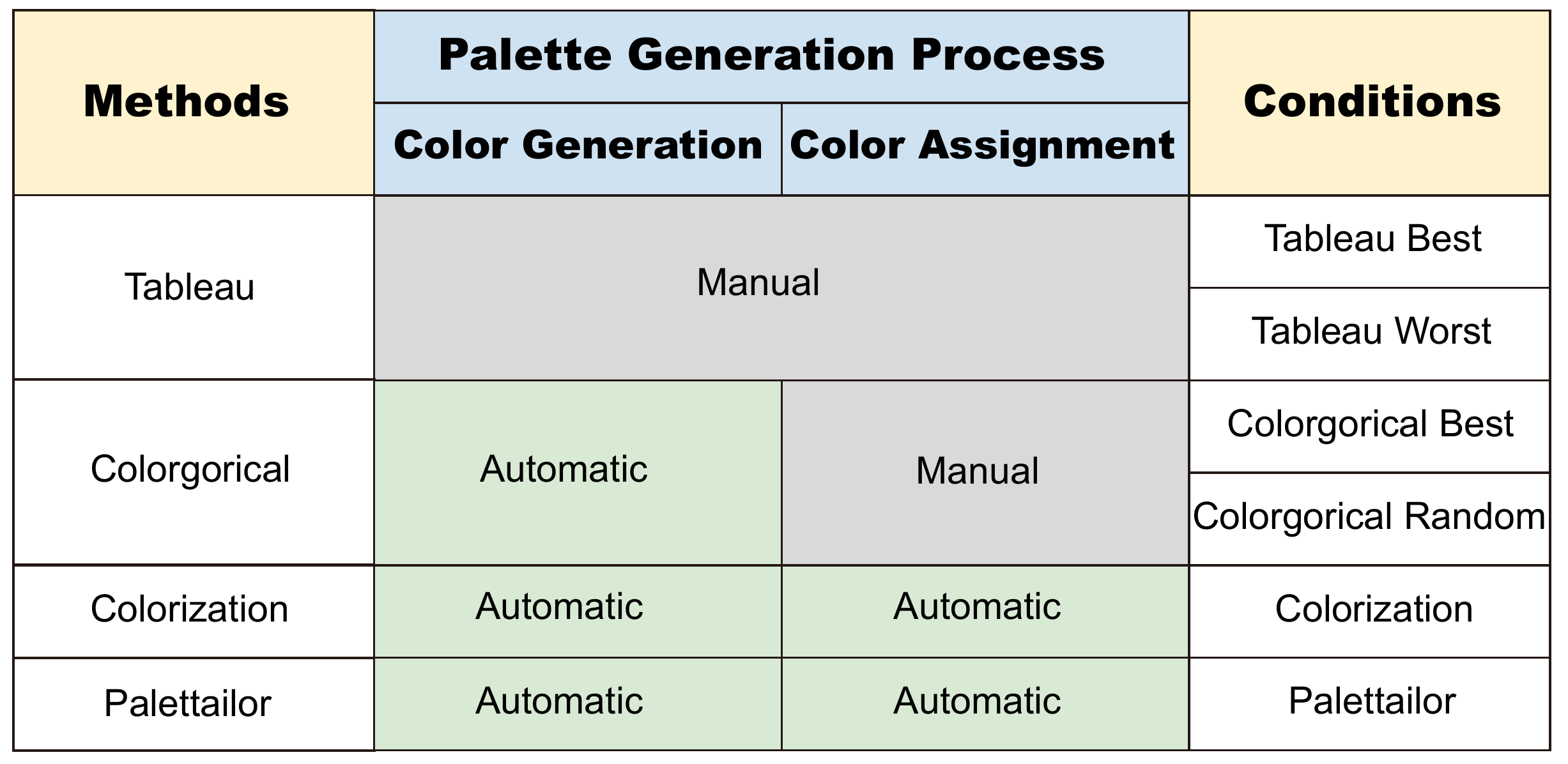}
	\caption{
		On the left column we show the existing methods that generate color palettes for visualizations including $Tableau$\cite{tableau}, $Colorgorical$\cite{gramazio2017colorgorical} and $Colorization$\cite{chen2014visual}, together with our proposed method $Palettailor$.
		These methods adopt different palette generation processes that have different levels of automation (shown in the middle columns).
		For example, $Colorgorical$'s process involves automatic color palette generation and manual color assignment, meaning that it needs manual effort to assign the colors from the palette to the visualization.
	}
	\label{fig:conditions}
	\vspace{-2mm}
\end{figure}

We evaluate the quality of our palette generation method by comparing the resulting visualizations from our method with those from the existing methods: $Tableau$\cite{tableau}, $Colorgorical$\cite{gramazio2017colorgorical} and $Colorization$\cite{chen2014visual}.
As shown in ~\autoref{fig:conditions}, the existing palette generation methods have different levels of automation.
For fully manual ($Tableau$) or partly manual methods ($Colorgorical$), we compared our method with both a best and a worst/random palette result (more details below).

To test these methods, we conducted two experiments  through Amazon Mechanical Turk (AMT) with 155 participants in total.
Experiment 1 focuses on multi-class scatterplots, while Experiment 2 tests our approach on line charts.
We test the different methods with two types of tasks:
discrimination tasks and preference tasks. In discrimination tasks, we ask participants questions that necessitate them to visually separate, or ``discriminate'', data from differently colored classes/lines. In preference tasks, we ask them about their subjective preferences between visual encodings with colors from the different palette generation methods.

\paragraph{Discrimination hypotheses.} For discrimination tasks, we had the following hypotheses:
\setitemize{noitemsep,topsep=0pt,parsep=0pt,partopsep=0pt}
\setenumerate{noitemsep,topsep=0pt,parsep=0pt,partopsep=0pt}
\begin{enumerate}[start=1,label={\bfseries H\arabic*}]
\item Palettailor's results are \yh{not worse than} the best cases from the manual process of designer-crafted palettes ($Tableau~Best$) and auto-generated palettes ($Colorgorical~Best$). 
\item Palettailor's results are better than the worst or random cases from the manual process of designer-crafted palettes ($Tableau~Worst$) and auto-generated palettes ($Colorgorical~Random$).
\item Palettailor's results are better than the existing fully-automated method ($Colorization$). 
\end{enumerate}

\vspace{2mm}
Besides discriminability, we also want to find out that whether our results are aesthetically preferred by people or not, thus we have the other two hypotheses:
\vspace{2mm}
\paragraph{Preference hypotheses.} In terms of preference tasks, we have the following hypotheses:
\begin{enumerate}[start=4,label={\bfseries H\arabic*}]
\item The preferences of Palettailor's results are \yh{not worse than} designer-crafted palettes ($Tableau$). 
\item Palettailor's results are preferred over the auto-generated palettes ($Colorgorical$ and $Colorization$). 
\end{enumerate}

\subsection{Experiment 1: Scatterplot Experiment}

In this experiment, each participant completed one of three tasks from two different discrimination tasks and one preference task.
As stimulus we used 30 multi-class scatterplot datasets for each task. For coloring the classes, we included six palette conditions -- one from our method and the other five from existing methods (see right column of ~\autoref{fig:conditions}).

\subsubsection{Experimental Design}
\vspace*{-5pt}
\paragraph{Tasks \& measures}.
We tested two discrimination tasks, and one preference task.
\begin{itemize}
  \item \emph{Counting (discrimination) task.}
Following the methodology by Wang et al.~\cite{wang2019optimizing}, we asked participants to count the number of classes in a scatterplot and to choose an answer among several options.
For each participant, we recorded the \emph{time} taken for each trial, and counted the \emph{errors} by calculating the differences between the actual number of classes and the participant's response. The counting task focuses on global discriminability.
We reported the \emph{error} measure as \emph{time} is not of primary interest for our hypotheses.
  \item \emph{Comparison (discrimination) task.}
  Following the methodology by Gleicher et al.~\cite{gleicher2013perception}, we asked participants to judge which of  two specified class centroids was higher 
  and to choose a value from a choice with two-alternatives.
  For each participant, we recorded the \emph{time} taken and \emph{error(0/1)} for each trial.    The comparison task focuses on local discriminability.
\item \emph{Preference task.}
Participants were shown a series of image pairs, each containing two scatterplots selected from two different conditions: one from $Palettailor$ and the other from a benchmark condition.
For each figure, the participant was asked to choose the plot they preferred.
A neutral choice button was also provided.
\end{itemize}
\yh{To avoid potential carryover effects between tasks, we measured preference on the colored visualizations as a stand-alone task, instead of measuring it in the context of each discrimination task.}

\vspace*{-5pt}
\paragraph{Conditions \& sample palette generation}.
As shown in ~\autoref{fig:conditions}, we have six conditions in total, one from our method ($Palettailor$), and the other five conditions from the existing methods ($Tableau~Best$, $Tableau~Worst$, $Colorgorical~Best$, $Colorgorical~Random$, and $Colorization$).

Since $Tableau$ is a designer-crafted method, we included two conditions to mimic the best and worst discriminable results generated from the manual process.
In a situation, for example, where the class number $m$ is smaller than 10, a designer needs to manually choose $m$ colors from the Tableau 10 palette.
We run  Wang et al.'s color assignment method \cite{wang2019optimizing} on the Tableau palette and select the first $m$ colors with the highest and lowest scores for generating what we call  $Tableau~Best$ and $Tableau~Worst$ conditions. We repeated this process multiple times and used the most suitable palettes as the final version in the experiment.

For Colorgorical, we specified the parameters in a way that we received a Low-Error setting:
20\% Perceptual Distance, 40\% Name Difference, 40\% Pair Preference (see original paper for more details \cite{gramazio2017colorgorical}).
We repeatedly generated 5 palettes with these settings via their tool and randomly chose one of them. We then applied the color assignment process from Wang et al. \cite{wang2019optimizing} to simulate a designer's optimized assignment, and used a random assignment to represent a common default result
($Colorgorical~Best$ and $Colorgorical~Random$ conditions).
We directly used Chen’s method~\cite{chen2014visual}  as the condition $Colorization$, since it is a fully-automated process. For Palettailor, the palettes were generated using the default weight setting of our method: [1,1,1].  Pilot tests showed that this setting performed best among different weight settings.

\paragraph{Scatterplot dataset generation}.
We used 30 multi-class scatterplot datasets with 6 to 10 classes in the experiment.
Each dataset had two target classes that were meant for comparison.
We generated the data points of the target classes using the method by Gleicher et al.~\cite{gleicher2013perception}.
The data points were placed according to a uniform random distribution in a $400\times400$ area, the height difference between these two classes was 30 pixels.
Additionally, we generated the rest of the classes (distractors) using Gaussian random sampling and randomly placed them in an $500\times500$ area.
This way the distractors have two kinds of relations to the target classes: intersecting them and including them.
We used the Gaussian process to control the overall properties of each class such as size and density~\cite{sedlmair2012taxonomy}.
To simplify the variety of visual shapes we defined sizes to be large or small, and densities to be sparse or dense.
Hence there were four types of classes: large \& dense ($n = 100$, $\sigma = 50$), small \& dense ($n = 50$, $\sigma = 20$), large \& sparse ($n = 50$, $\sigma = 100$), and small \& sparse ($n = 20$, $\sigma = 50$). See the supplemental material for more details.

\begin{table}[tbp]
\centering
\small
\caption{Organization of Datasets for the Discrimination Tasks: 30 datasets $\times$ 6 conditions. C: condition; G: participant group.}
\begin{tabular}{r | c c c c c c}
\hline
$$ & C1 & C2 & C3 & C4 & C5 & C6 \\
 \hline
\textit{Dataset 1} & \textbf{G1} & G2  & G3 & G4 & G5  & G6  \\
\textit{Dataset 2} & G6 & \textbf{G1}  & G2 & G3 & G4  & G5  \\
\textit{Dataset 3} & G5 & G6  & \textbf{G1} & G2 & G3  & G4  \\
\textit{Dataset 4} & G4 & G5  & G6 & \textbf{G1} & G2  & G3  \\
\textit{Dataset 5} & G3 & G4  & G5 & G6 & \textbf{G1}  & G2  \\
\textit{Dataset 6} & G2 & G3  & G4 & G5 & G6  & \textbf{G1}  \\
\textit{...} &   &    &   &   &    &    \\
\textit{Dataset 29} & G3 & G4  & G5 & G6 & \textbf{G1}  & G2  \\
\textit{Dataset 30} & G2 & G3  & G4 & G5 & G6  & \textbf{G1}  \\
\hline
\end{tabular}
\label{tab:conditions}
\end{table}
%

\vspace*{-5pt}
\paragraph{Experiment organization}.
We used a \emph{between-subjects} layout in the experiment to test the effects of 6 conditions using 30 datasets. We did not adopt a within-subjects design (i.e., each participant going through all the 180 condition-dataset combinations) because this would lead to unwanted learning effects, as one would see each dataset repeatedly six times.
Thus instead, we had each participant to see different subsets of datasets for different conditions.
As shown in Table \ref{tab:conditions}, we used a Latin Square  arrangement for the combination of datasets and conditions to avoid ordering effects.
Thus there were six participant groups, each having 30 trials, every participant was assigned to one of the groups.

%

\vspace*{-5pt}
\paragraph{Pilot study \& power analysis}.
We conducted a pilot study to check the experimental setup.
It was also used to infer first insights about the effect sizes to be expected, the effect sizes were fed  into a power analysis.
For example, the power analysis
 (with an effect size Cohen’s d of 0.4, alpha of 0.05 and beta level of 0.8)
 suggested a minimum number of 24 participants
for detecting a meaningful difference between $Palettailor$ and $Colorgorical~Best$ in the counting task.
\begin{table}[tbp]
\centering
\small
\caption{
 Participants recruited for the formal study. G\# specifies the participant group (more details see text) }
\begin{tabular}{r | p{0.15\linewidth}<{\centering} p{0.05\linewidth}<{\centering} p{0.05\linewidth}<{\centering} p{0.05\linewidth}<{\centering} p{0.05\linewidth}<{\centering} p{0.05\linewidth}<{\centering} p{0.05\linewidth}<{\centering}}
\hline
\textbf{Task} & \textbf{Participants} & \textbf{G1} & \textbf{G2} & \textbf{G3} & \textbf{G4} & \textbf{G5} & \textbf{G6} \\
 \hline
\textit{Counting} & 47 & 9 & 9 & 9 & 7 & 6 & 7 \\
\textit{Comparison} & 33 & 6 & 5 & 6 & 5 & 5 & 6 \\
\textit{Preference-1} & 25 & 5 & 5 & 5 & 5 & 5 & N\/A \\
\textit{Slope} & 30 & 8 & 7 & 5 & 5 & 5 & N\/A \\
\textit{Preference-2} & 20 & 5 & 5 & 5 & 5 & N\/A & N\/A \\
\hline
\end{tabular}
\label{tab:formalResults}
\vspace{-4mm}
\end{table}

\vspace*{-5pt}
\paragraph{Participants}.
We recruited 105 participants for the main experiment using a crowd sourcing platform.
According to the completion times in the pilot study, we paid each \$2.5 for the Counting Task, \$1.5 for the Comparison Task and \$0.5 compensation for Preference Task,  in order to exceed the US minimum wage.
All the participants gave us their informed consent, and no participant reported color vision deficiency.

\vspace*{-5pt}
\paragraph{Procedure}.
Each participant went through the following procedure:
(1) viewing the instructions of the task and completing two training trials (except Preference Task);
(2) performing the task as accurately as possible based on the instruction to do so;
and (3) providing basic demographic information. See the supplemental material for more details.

\begin{figure*}[ht]
\centering
\includegraphics[width=\linewidth]{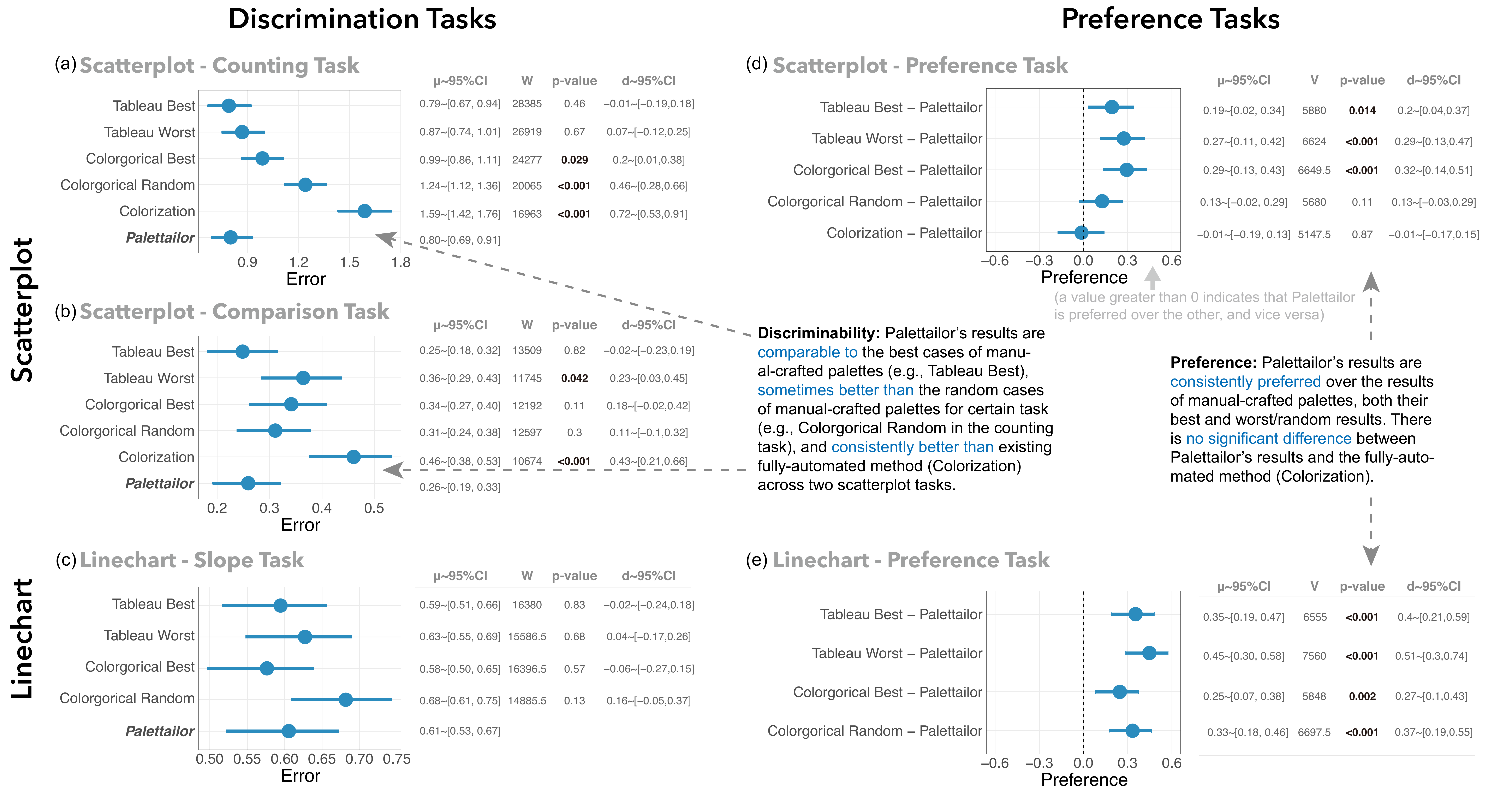}
\caption{
Results for scatterplot and line chart experiments:
For each task (e.g., the counting task in the scatterplot experiment, marked as \emph{Scatterplot - Counting Task}), we give a confidence interval plot and a statistic table.
In the table, for each condition we provide the statistics including the mean with 95\% confidence interval ($\mu\sim$95\%CI), the W-value and p-value from the Mann-Whitney test, and the effect size ($d\sim$95\%CI). For the discrimination tasks on the left, we conducted statistical tests to compare $Palettailor$ to every other condition, and provide W or V values, p-value and effect size ($d\sim$95\%CI) accordingly in the tables.
}
\label{fig:taskResults}
\end{figure*}

\subsubsection{Results}
Following the methodology of previous  studies, we analyzed the results using 95\% confidence intervals, and also conducted Mann-Whitney tests to compare the differences between conditions. We selected this more conservative, non-parametric test as we observed in the pilot study that some measures were non-normally distributed.
We also computed the effect size using \emph{Cohen's d}, i.e., the difference in means of the conditions divided by the pooled standard deviation.
In addition, we checked the data collected for both pilot and main studies, and found no significant interaction effects between datasets and conditions, so we focus on reporting the main effects.

\vspace*{-5pt}
\paragraph{Discriminability}.
~\autoref{fig:taskResults}(a)\&(b) show the error results for the counting and comparison tasks, including effect sizes and p-values.
Here, we focus on the analysis of \emph{error}. We also measured \emph{time}, but as it is not of primary interest for our hypotheses we provide the results in the supplemental materials.

The results are generally aligned with our discriminability hypotheses  (\textbf{H1-3}).

First, we compared $Palettailor$ to the best cases of the manual processes, i.e., $Tableau~Best$ and $Colorgorical~Best$.
We found that the $Palettailor$ condition has significantly less errors ($p = 0.029$) than $Colorgorical~Best$ in the counting task.
We did not observe other significant differences between these conditions.
The result indicates that our approach \yh{appears not to be worse than} the best cases of manually designer-crafted palettes and auto-generated palettes (\textbf{H1}), and might even lead to a better performance in the counting task.

Second, we compared $Palettailor$ to the random or worst cases of the manual processes, i.e., $Tableau~Worst$ and $Colorgorical~Random$.
We found that in the counting task, $Palettailor$ has a significantly lower error rate ($p < 0.001$) than $Colorgorical~Random$. In the comparison task, $Palettailor$ has a significant lower error rate ($p = 0.042$) than $Tableau~Worst$.
This partially confirms \textbf{H2} that our method is better than some of the manually created worst or random cases for certain discrimination tasks.

Finally, $Palettailor$ is clearly better than the $Colorization$ method, for both the counting task and the comparison task ($p < 0.001$).
Thus, \textbf{H3} is fully confirmed and our method is indeed better than the existing fully-automated method in terms of discriminability.

\vspace*{-5pt}
\paragraph{Preference}.
For each comparison within the preference experiment, we assigned a preference score, i.e., $1$ for preferring $Palettailor$, $-1$ for preferring another condition, and $0$ for a neutral choice.
We aggregated the scores for each pair containing $Palettailor$ and a benchmark condition, e.g., \emph{Tableau Best - Palettailor}.
An average score value greater than 0 indicates that $Palettailor$ is preferred over the other condition, and vice versa.
The results are summarized in ~\autoref{fig:taskResults}(d).

We found that $Palettailor$ is significantly preferred over the designer-crafted palettes: $Tableau~Best$ ($p = 0.014$) and $Tableau~Worst$ ($p < 0.001$).
This finding is not aligned with our hypothesis \textbf{H4}.
We also found that $Palettailor$ is significantly preferred over $Colorgorical~Best$ ($p < 0.001$), yet there is no clear preference between $Palettailor$ and $Colorgorical~Random$ or $Colorization$.
This observation partially aligns with \textbf{H5} and will be further discussed in Section \ref{sec:discussion}.

\subsection{Experiment 2: Line Chart Experiment}

In addition to scatterplots, we also applied our method to line charts and evaluated its effectiveness in Experiment 2.
We used 30 line chart datasets for each task as stimulus.
The setup of Experiment 2 was similar to Experiment 1; the differences between the two will be described below.

\subsubsection{Experimental Design}


\vspace*{-5pt}
\paragraph{Tasks \& measures}.
Each participant completed one of two tasks:
\begin{itemize}
\item \emph{Slope (discrimination) task. }
  Following the methodology by Javed et al.~\cite{javed2010graphical}, we asked participants to find out which line had the largest increase
  and to choose an answer among several options. For each participant, we recorded the \emph{error (0/1)} and \emph{time} taken for each trial. Similar to Experiment 1, we focused on reporting the \emph{error} measure, see the supplemental materials for the details on \emph{time}.
\item \emph{Preference task.}
Similar to Experiment 1, each participant was shown a series of image pairs, each containing two line charts, one from the $Palettailor$ condition and the other from one of the benchmark conditions.
The participant was asked to choose the one s/he preferred.
\end{itemize}

\vspace*{-5pt}
\paragraph{Conditions \& sample palettes}.
In contrast to Experiment 1, we only tested five conditions in total, our method ($Palettailor$), as well as $Tableau~Best$, $Tableau~Worst$, $Colorgorical~Best$, and $Colorgorical~Random$.
We did not include $Colorization$ as it is not applicable to line charts.
We used the same process as in Experiment 1 to generate the sample palettes for every condition.

\vspace*{-5pt}
\paragraph{Line chart dataset generation}.
We generated 30 different line charts with 6-10 lines following a carefully designed method.
Every line chart had a \emph{target} line with the largest increase; other lines with smaller increases were used to distract participants.
We generated the target line with an increase of 0.35, with every point being within $[k-0.175,~k+0.175]$ ($k$ is a random number and $k \in [0,1]$).
Then we repeated the process to generate the rest of the lines, each with an increase less than 0.3, i.e., every point within $[k-0.15,k+0.15]$.
Each line chart was displayed in an $400\times400$ area, thus the minimum difference of increase between the largest and other lines on the screen is $0.05\times400=20$ pixels. See the supplemental material for more details.

The other aspects of the experiment were similar to Experiment 1.
We used a \emph{between-subjects} layout to avoid unwanted learning effects.
We had 30 datasets and 5 conditions, the participants were divided into five groups, each going through a subset of 30 condition-dataset combinations.
We conducted a pilot study including 11 participants from a crowd sourcing platform.
As shown in Table \ref{tab:formalResults}, we recruited 50 participants in total for Experiment 2.
Each participant went through the same procedure as that in Experiment 1, and was paid \$1.5 for the Slope Task and \$0.5 compensation for the Preference Task to exceed the US minimum wage.

\subsubsection{Results}

\vspace*{-5pt}
\paragraph{Discriminability}.
~\autoref{fig:taskResults}(c) shows the error results for the slope task.
We observed no significant differences between $Palettailor$ and $Tableau~Best$ or $Colorgorical~Best$, which indicates that our approach is \yh{not worse than} the best cases of manually  designer-crafted palettes and auto-generated palettes (\textbf{H1}).
Also, we did not observe clear differences between $Palettailor$ and the worst or random manual cases, i.e., $Tableau~Worst$ or $Colorgorical~Random$.
This finding does not aligned with \textbf{H2}.
We discuss the detailed implications of these results in the next section.

\vspace*{-5pt}
\paragraph{Preference}.
The results are summarized in ~\autoref{fig:taskResults}(e), where an average preference score larger than 0 means that $Palettailor$ was preferred over the other condition, and vice versa.
We found that $Palettailor$ is preferred over the designer-crafted palettes: $Tableau~Best$ ($p < 0.001$) and $Tableau~Worst$ ($p < 0.001$).
This finding does not align with our hypothesis \textbf{H4}.
We also found that $Palettailor$ is preferred over $Colorgorical~Best$ ($p = 0.002$) and $Colorgorical~Random$ ($p < 0.001$), which aligns with \textbf{H5}.

\begin{table}[tbp]
\centering
\small
\caption{
 The summary of results. E1 and E2 indicate the scatterplot and line chart experiment. (**) means a hypothesis is confirmed; (*) means partly confirmed.}
\begin{tabular}{l | cc}
\hline
Hypothesis & \textit{E1} & \textit{E2}\\
 \hline
\textbf{H1} Discriminability is \yh{not worse than} the best manual cases. & ** & **  \\
\textbf{H2} Discriminability is better than the random manual cases. & * &  \\
\textbf{H3} Discriminability is better than the automated method. & ** & NA \\
\textbf{H4} Preference is \yh{not worse than} the designer-crafted palettes. &  &  \\
\textbf{H5} Preference is better than the auto-generated palettes. & * & **  \\
\hline
\end{tabular}
\label{tab:hypotheses}
\vspace{-4mm}
\end{table}

\subsection{Discussion}
\label{sec:discussion}

The results of the two experiments are summarized in Table \ref{tab:hypotheses}.
In terms of \emph{discriminability}, we found that Palettailor \yh{appears not to be worse than} the best cases of color generation processes that involve manual steps (i.e., the $Tableau~Best$ and $Colorgorical~Best$ conditions).
We also found that Palettailor is more effective than some less-optimized manual cases for certain tasks (e.g., the $Colorgorical~Random$ condition in the scatterplot-counting task).
Furthermore, Palettailor is consistently better than the fully-automated approach (i.e., $Colorization$).
The results indicate that Palettailor can automatically generate palettes that are \textbf{\textit{\yh{not worse than}}} those from the manual approaches, and that are \textbf{\textit{more discriminable}} than those from the fully-automated palette generation approach.

However, although the results regarding discriminability are generally aligned with our hypotheses for the scatterplot experiment, we found no significant differences in the line chart experiment.
One possible explanation is that the visual complexity of a line chart is generally lower than for a scatterplot.
Thus, having discriminable clusters in a scatterplot may aid more to people's visual performance than having discriminable lines in a line chart.
This leads to the necessity  to further explore how Palettailor can benefit visual perception when applying it to different visualization types.

In terms of \emph{preference}, we observed that Palettailor is preferred over designer-crafted palettes (i.e., the $Tableau$ palettes), and that it is also preferred over some auto-generated palettes for certain visualizations (e.g., the $Colorgorical$ palettes for scatterplot visualizations).

From our results  it seems like that visualizations with brighter colors were generally preferred (i.e., the $Palettailor$, $Colorization$ and sometimes $Colorgorical$ palettes), and yet the designer-crafted palette $Tableau$ was less preferred.
There are several potential explanations.
It is possible that the preference of bright colors is due to the  cultural background of the participants~\cite{palmer2010ecological}, given that all the participants were recruited from the US.
Another explanation is that for some people, when looking at visualizations with bright colors, they may gain \emph{cognitive ease}~\cite{tversky1974judgment}.
In other words, they may be biased towards \emph{believing} that it would be easier to tell apart clusters or  lines (i.e., performing discrimination tasks), no matter whether the colors could truly aid  discriminability.
\yh{The result can also potentially be explained by \emph{affective response} \cite{bartram2017affective}, i.e., more saturated colors are strongly correlated with intensity and positive impressions. Such affect might have caused higher rates of the brighter palettes.}
Future research may benefit the design of Palettailor in supporting people's aesthetic preferences.



\yh{
\subsection{Limitations}

The goal of our user study was an initial exploration of the effectiveness of our method.
We thus focused on benchmarks from systematically generated datasets and sample palettes.
At the same time, we fixed or limited some of the other factors, which may affect the generalizability of the study results.
Although our proposed method is capable of supporting a variety of visualization types, in the study we only chose two types: scatterplots and line charts.
While these are common visualization types, we did not cover visualizations with larger area marks, such as bar or area charts.
In addition, our method is capable of tailoring  palettes based on the background color, but for the study, we chose a white background as it is the most commonly used canvas color to display visualizations.
Lastly, in our study we measured aesthetic preferences as a stand-alone task to avoid potential carryover effects. However, it might be possible that people's preference would change based on the tasks they are doing. Further studies need to be conducted to comprehensively examine these factors.

} 
\section{Interactive System for Colorization}

\label{sec:interaction}
\begin{figure}[t]
\centering
\includegraphics[width=\linewidth]{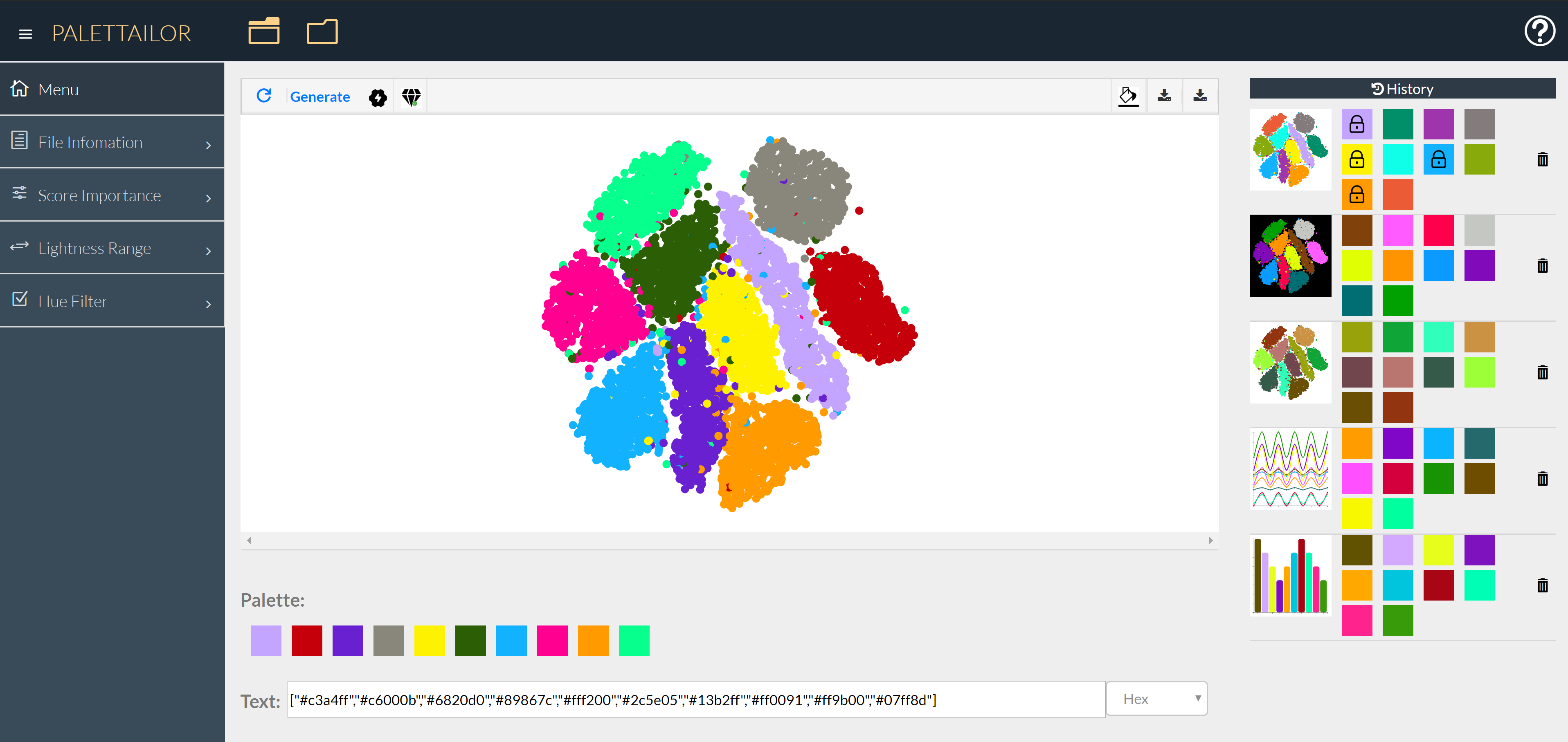}
\caption{A screenshot of our interactive system. Users can load different datasets and generate suitable color palettes. If they are satisfied with a result, they can download the image or export the palette definition to use in their own system. We also offer a history for users to find their previous results.}
\label{fig:palettailor}
\vspace{-4mm}
\end{figure}

Based on the automatic approach presented above, 
we developed an interactive color palette generation system (Fig.~\ref{fig:palettailor}) that allows users to feed in personal preferences when looking for good colors. 
After the user uploads his/her data,  different color palettes will be produced automatically. Users may also edit color palettes based on their preferences.
A history and other interactive features support the user in finding a desired color palette.

\subsection{User Interaction}
\label{sec:constraint}
Our system provides three main ways to interact with color palettes: 
adjustable palette generation, constrained palette generation, and palette completion.

\paragraph{Adjustable palette generation}.
Users can adjust the weights of the three scoring functions to generate palettes with different emphasis.
For example, if a user wants to emphasize on local discrimination, then s/he can turn up  Point Distinctness or turn down Color Discrimination or Name Difference. If a user wants to create a palette composed of as many color names as possible, s/he can turn up Name Difference.

\paragraph{Constrained palette generation}.
Users can also specify preferable colors for palette generation. We use 11 universal basic color terms (blue, brown, green, orange, pink, purple, red, yellow, black, grey, and white)~\cite{berlin1991basic} as our Hue Filter.
For example, when a user selects some colors for generating a color palette, our system will automatically search for optimized palettes within the respective ranges.
As our algorithm takes the background color into account, we also allow users to specify a background color and adjust the luminance range accordingly.

\paragraph{Palette completion}.
Our system enables users to fix colors of a palette. 
Once some colors are fixed, our system will evaluate the current palette as well as the data characteristics, and search for optimized other colors to complete it.

\begin{figure}[h]
\centering
\includegraphics[width=\linewidth]{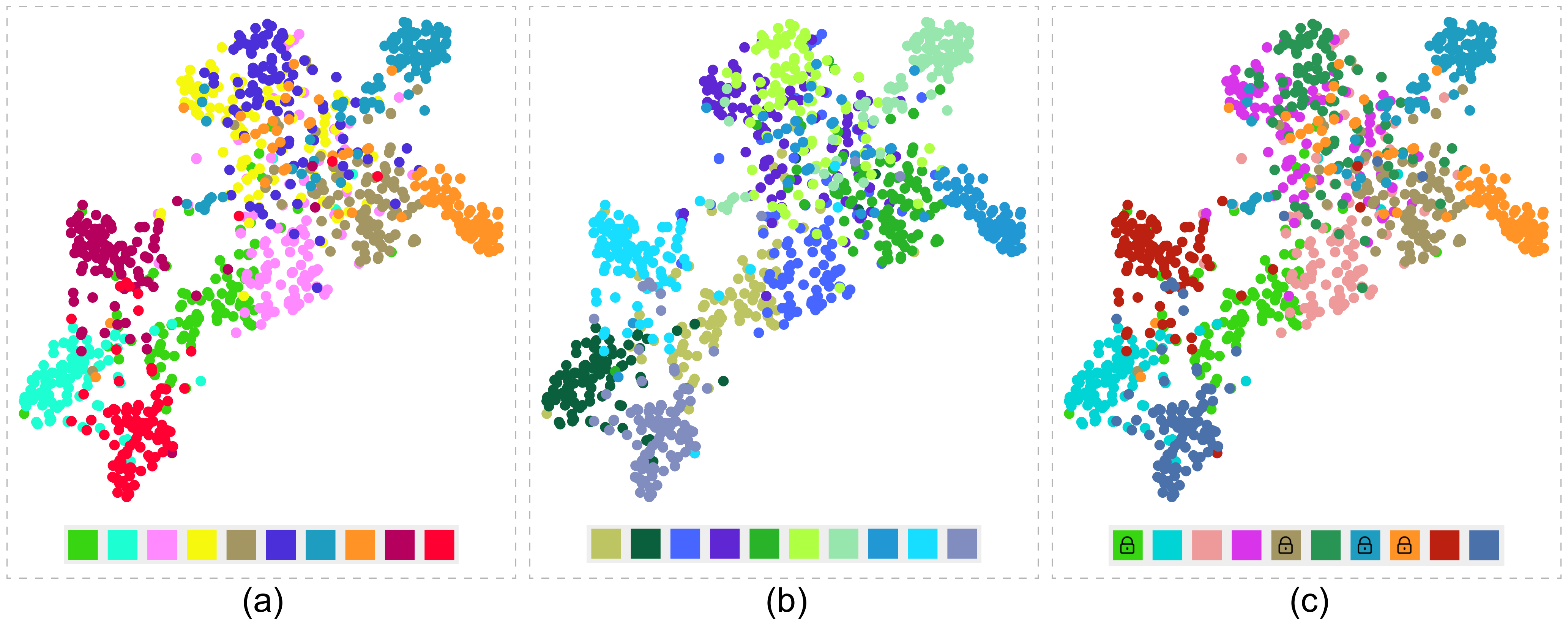}
\caption{CIFAR10 dataset: (a) Palette generated by default setting; (b) User-specified Hue Filter with ``green" and ``blue"; (c) After clicking on colors to lock them, our system completes the palette automatically.}
\label{fig:casestudy}
\vspace{-4mm}
\end{figure}

\subsection{Case Study}

We conducted a case study with a data expert to evaluate the usability of our system.  The dataset we used for this study was the CIFAR10 dataset~\cite{krizhevsky2009learning} that includes classification results for images of ten object classes (e.g., cats, ships).
The expert first used the default settings of our system to automatically produce a palette for assigning colors to different objects in the dataset.
Seeing the result in Fig.~\ref{fig:casestudy}(a), she was able to easily discriminate every class.
After generating multiple palettes with our default settings, she used the Hue Filter and chose two preferable colors (``green" and ``blue"). The result is shown in  Fig.~\ref{fig:casestudy}(b). Even though all colors are variants of blue and green, it is still possible to discriminate all classes.
However, the expert preferred the previous results by our default setting, so she used the history to find the results. Rather than generating a totally new palette, she wanted to preserve some colors. So she clicked on these colors to lock them and produced another palette, shown in Fig.~\ref{fig:casestudy}(c). Overall, she was very satisfied with the system and used all our modalities to create a good palette.

\section{Conclusion}
We presented Palettailor, a data-aware approach for producing color palettes for categorical visualizations that allows a better visual discrimination of classes while being visually pleasing. This goal is achieved by a simulated annealing-based optimization, which we feed with three scoring functions: point distinctness, name difference, and color discrimination constraint.
We evaluated Palettailor through a crowd-sourcing study, which empirically demonstrated that our produced palettes allow for good class discrimination, are preferred by users, and can do that fully automatically.

Our current measure of point distinctness is based on position and color of the points in the underlying visualizations. However, class discriminability in human perception is also related to many other factors such as mark shape, mark size, and rendering order~\cite{micallef2017towards}. In the future, we would like to extend our model to incorporate such factors as well. \yh{Weighting the point distinctness by the number of points in each class is able to generate reasonable results in most cases. However, it might de-emphasize classes with fewer data points, such as outliers. To address this issue, we will integrate a user-steered parameter so as to assign  priorities to small categories.}
In addition, we would also like to let our technique support more visualization techniques such as pie charts and parallel coordinates. We also want to include the generation of color palettes  for people with color vision deficiencies.

\acknowledgments{
This work is supported by the grants of the NSFC (61772315, 61861136012), the Open Project Program of State Key Laboratory of
Virtual Reality Technology and Systems, Beihang University (No.VRLAB2020C08), the CAS grant (GJHZ1862) and
Deutsche Forschungsgemeinschaft (DFG) -- Project-IDs DE 620/26-1, as well as 251654672 -- TRR 161 Quantitative methods for visual computing. }

\bibliographystyle{abbrv-doi}


\begin{thebibliography}{10}

\bibitem{aarts1989stochastic}
E.~Aarts.
\newblock A stochastic approach to combinatorial optimization and neural
  computing.
\newblock {\em Simulated Annealing and Boltzmann Machines}, 1989.

\bibitem{aupetit2016sepme}
M.~Aupetit and M.~Sedlmair.
\newblock {SepMe}: 2002 new visual separation measures.
\newblock In {\em IEEE Pacific Visualization Symposium}, pp. 1--8, 2016. doi:
  {{%
10\hspace{.1pt}\discretionary{.}{%
}{.}\hspace{.4pt}1109\discretionary{/}{%
}{/}pacificvis\hspace{.1pt}\discretionary{.}{%
}{.}\hspace{.4pt}2016\hspace{.1pt}\discretionary{.}{%
}{.}\hspace{.4pt}7465244}}


\bibitem{bartram2017affective}
L.~Bartram, A.~Patra, and M.~Stone.
\newblock Affective color in visualization.
\newblock In {\em Proceedings of the 2017 CHI conference on human factors in
  computing systems}, pp. 1364--1374, 2017.

\bibitem{berlin1991basic}
B.~Berlin and P.~Kay.
\newblock {\em Basic color terms: Their universality and evolution}.
\newblock Univ of California Press, 1991.

\bibitem{brychtova2017effect}
A.~Brychtov{\'a} and A.~{\c{C}}{\"o}ltekin.
\newblock The effect of spatial distance on the discriminability of colors in
  maps.
\newblock {\em Cartography and Geographic Information Science}, 44(3):229--245,
  2017.

\bibitem{chen2014visual}
H.~Chen, W.~Chen, H.~Mei, Z.~Liu, K.~Zhou, W.~Chen, W.~Gu, and K.-L. Ma.
\newblock Visual abstraction and exploration of multi-class scatterplots.
\newblock {\em IEEE Trans. Vis. \& Comp. Graphics}, 20(12):1683--1692, 2014.

\bibitem{chuang2009energy}
J.~Chuang, D.~Weiskopf, and T.~M{\"o}ller.
\newblock Energy aware color sets.
\newblock {\em Computer Graphics Forum}, 28(2):203--211, 2009. doi: {{%
10\hspace{.1pt}\discretionary{.}{%
}{.}\hspace{.4pt}1111\discretionary{/}{%
}{/}j\hspace{.1pt}\discretionary{.}{%
}{.}\hspace{.4pt}1467\discretionary{%
}{-}{-}8659\hspace{.1pt}\discretionary{.}{%
}{.}\hspace{.4pt}2009\hspace{.1pt}\discretionary{.}{%
}{.}\hspace{.4pt}01359\hspace{.1pt}\discretionary{.}{%
}{.}\hspace{.4pt}x}}


\bibitem{fang2017categorical}
H.~Fang, S.~Walton, E.~Delahaye, J.~Harris, D.~Storchak, and M.~Chen.
\newblock Categorical colormap optimization with visualization case studies.
\newblock {\em IEEE Trans. Vis. \& Comp. Graphics}, 23(1):871--880, 2017. doi:
  {{%
10\hspace{.1pt}\discretionary{.}{%
}{.}\hspace{.4pt}1109\discretionary{/}{%
}{/}tvcg\hspace{.1pt}\discretionary{.}{%
}{.}\hspace{.4pt}2016\hspace{.1pt}\discretionary{.}{%
}{.}\hspace{.4pt}2599214}}


\bibitem{gleicher2013perception}
M.~Gleicher, M.~Correll, C.~Nothelfer, and S.~Franconeri.
\newblock Perception of average value in multiclass scatterplots.
\newblock {\em IEEE Trans. Vis. \& Comp. Graphics}, 19(12):2316--2325, 2013.
  doi: {{%
10\hspace{.1pt}\discretionary{.}{%
}{.}\hspace{.4pt}1109\discretionary{/}{%
}{/}tvcg\hspace{.1pt}\discretionary{.}{%
}{.}\hspace{.4pt}2013\hspace{.1pt}\discretionary{.}{%
}{.}\hspace{.4pt}183}}


\bibitem{gramazio2017colorgorical}
C.~C. Gramazio, D.~H. Laidlaw, and K.~B. Schloss.
\newblock Colorgorical: Creating discriminable and preferable color palettes
  for information visualization.
\newblock {\em IEEE Trans. Vis. \& Comp. Graphics}, 23(1):521--530, 2017. doi:
  {{%
10\hspace{.1pt}\discretionary{.}{%
}{.}\hspace{.4pt}1109\discretionary{/}{%
}{/}tvcg\hspace{.1pt}\discretionary{.}{%
}{.}\hspace{.4pt}2016\hspace{.1pt}\discretionary{.}{%
}{.}\hspace{.4pt}2598918}}


\bibitem{harrower2003colorbrewer}
M.~Harrower and C.~A. Brewer.
\newblock {ColorBrewer}.org: an online tool for selecting colour schemes for
  maps.
\newblock {\em The Cartographic Journal}, 40(1):27--37, 2003. doi: {{%
10\hspace{.1pt}\discretionary{.}{%
}{.}\hspace{.4pt}4324\discretionary{/}{%
}{/}9781351191234\discretionary{%
}{-}{-}18}}


\bibitem{healey1996choosing}
C.~G. Healey.
\newblock Choosing effective colours for data visualization.
\newblock In {\em Proc. IEEE Conf. on Visualization}, pp. 263--270, 1996. doi:
  {{%
10\hspace{.1pt}\discretionary{.}{%
}{.}\hspace{.4pt}1109\discretionary{/}{%
}{/}visual\hspace{.1pt}\discretionary{.}{%
}{.}\hspace{.4pt}1996\hspace{.1pt}\discretionary{.}{%
}{.}\hspace{.4pt}568118}}


\bibitem{heer2012color}
J.~Heer and M.~Stone.
\newblock Color naming models for color selection, image editing and palette
  design.
\newblock In {\em Proc. SIGCHI Conference on Human Factors in Computing
  Systems}, pp. 1007--1016, 2012. doi: {{%
10\hspace{.1pt}\discretionary{.}{%
}{.}\hspace{.4pt}1145\discretionary{/}{%
}{/}2207676\hspace{.1pt}\discretionary{.}{%
}{.}\hspace{.4pt}2208547}}


\bibitem{javed2010graphical}
W.~Javed, B.~McDonnel, and N.~Elmqvist.
\newblock Graphical perception of multiple time series.
\newblock {\em IEEE Trans. Vis. \& Comp. Graphics}, 16(6):927--934, 2010.

\bibitem{kirkpatrick1983optimization}
S.~Kirkpatrick, C.~D. Gelatt, M.~P. Vecchi, et~al.
\newblock Optimization by simmulated annealing.
\newblock {\em Science}, 220(4598):671--680, 1983.

\bibitem{krizhevsky2009learning}
A.~Krizhevsky and G.~Hinton.
\newblock Learning multiple layers of features from tiny images.
\newblock Technical report, Citeseer, 2009.

\bibitem{lee2013perceptually}
S.~Lee, M.~Sips, and H.-P. Seidel.
\newblock Perceptually driven visibility optimization for categorical data
  visualization.
\newblock {\em IEEE Trans. Vis. \& Comp. Graphics}, 19(10):1746--1757, 2013.
  doi: {{%
10\hspace{.1pt}\discretionary{.}{%
}{.}\hspace{.4pt}1109\discretionary{/}{%
}{/}tvcg\hspace{.1pt}\discretionary{.}{%
}{.}\hspace{.4pt}2012\hspace{.1pt}\discretionary{.}{%
}{.}\hspace{.4pt}315}}


\bibitem{lin2013selecting}
S.~Lin, J.~Fortuna, C.~Kulkarni, M.~Stone, and J.~Heer.
\newblock Selecting semantically-resonant colors for data visualization.
\newblock {\em Computer Graphics Forum}, 32(3pt4):401--410, 2013. doi: {{%
10\hspace{.1pt}\discretionary{.}{%
}{.}\hspace{.4pt}1111\discretionary{/}{%
}{/}cgf\hspace{.1pt}\discretionary{.}{%
}{.}\hspace{.4pt}12127}}


\bibitem{machado2009physiologically}
G.~M. Machado, M.~M. Oliveira, and L.~A. Fernandes.
\newblock A physiologically-based model for simulation of color vision
  deficiency.
\newblock {\em IEEE Trans. Vis. \& Comp. Graphics}, 15(6):1291--1298, 2009.
  doi: {{%
10\hspace{.1pt}\discretionary{.}{%
}{.}\hspace{.4pt}1109\discretionary{/}{%
}{/}TVCG\hspace{.1pt}\discretionary{.}{%
}{.}\hspace{.4pt}2009\hspace{.1pt}\discretionary{.}{%
}{.}\hspace{.4pt}113}}


\bibitem{maxwell2000visualizing}
B.~A. Maxwell.
\newblock Visualizing geographic classifications using color.
\newblock {\em The Cartographic Journal}, 37(2):93--99, 2000. doi: {{%
10\hspace{.1pt}\discretionary{.}{%
}{.}\hspace{.4pt}1179\discretionary{/}{%
}{/}caj\hspace{.1pt}\discretionary{.}{%
}{.}\hspace{.4pt}2000\hspace{.1pt}\discretionary{.}{%
}{.}\hspace{.4pt}37\hspace{.1pt}\discretionary{.}{%
}{.}\hspace{.4pt}2\hspace{.1pt}\discretionary{.}{%
}{.}\hspace{.4pt}93}}


\bibitem{micallef2017towards}
L.~Micallef, G.~Palmas, A.~Oulasvirta, and T.~Weinkauf.
\newblock Towards perceptual optimization of the visual design of scatterplots.
\newblock {\em IEEE Trans. Vis. \& Comp. Graphics}, 23(6):1588--1599, 2017.
  doi: {{%
10\hspace{.1pt}\discretionary{.}{%
}{.}\hspace{.4pt}1109\discretionary{/}{%
}{/}TVCG\hspace{.1pt}\discretionary{.}{%
}{.}\hspace{.4pt}2017\hspace{.1pt}\discretionary{.}{%
}{.}\hspace{.4pt}2674978}}


\bibitem{Mittelstadt2015Color-32461}
S.~Mittelst\"{a}dt, D.~J\"{a}ckle, F.~Stoffel, and D.~A. Keim.
\newblock Colorcat : Guided design of colormaps for combined analysis tasks.
\newblock In E.~Bertini, ed., {\em Eurographics Conference on Visualization
  (EuroVis) : Short Papers}, pp. 115--119. The Eurographics Association, 2015.
  doi: {{%
10\hspace{.1pt}\discretionary{.}{%
}{.}\hspace{.4pt}2312\discretionary{/}{%
}{/}eurovisshort\hspace{.1pt}\discretionary{.}{%
}{.}\hspace{.4pt}20151135}}


\bibitem{Munzner:2001992}
T.~Munzner.
\newblock {\em Visualization Analysis and Design}.
\newblock AK Peters visualization series. CRC Press, Boca Raton, FL, 2015. doi:
  {{%
10\hspace{.1pt}\discretionary{.}{%
}{.}\hspace{.4pt}1109\discretionary{/}{%
}{/}pacificvis\hspace{.1pt}\discretionary{.}{%
}{.}\hspace{.4pt}2016\hspace{.1pt}\discretionary{.}{%
}{.}\hspace{.4pt}7465242}}


\bibitem{nardini2019making}
P.~Nardini, M.~Chen, F.~Samsel, R.~Bujack, M.~Bottinger, and G.~Scheuermann.
\newblock The making of continuous colormaps.
\newblock {\em IEEE Trans. Vis. \& Comp. Graphics}, December 2019. doi: {{%
10\hspace{.1pt}\discretionary{.}{%
}{.}\hspace{.4pt}1109\discretionary{/}{%
}{/}tvcg\hspace{.1pt}\discretionary{.}{%
}{.}\hspace{.4pt}2019\hspace{.1pt}\discretionary{.}{%
}{.}\hspace{.4pt}2961674}}


\bibitem{palmer2010ecological}
S.~E. Palmer and K.~B. Schloss.
\newblock An ecological valence theory of human color preference.
\newblock {\em Proceedings of the National Academy of Sciences},
  107(19):8877--8882, 2010.

\bibitem{pham2012intelligent}
D.~Pham and D.~Karaboga.
\newblock {\em Intelligent optimisation techniques: genetic algorithms, tabu
  search, simulated annealing and neural networks}.
\newblock Springer Science \& Business Media, 2012.

\bibitem{sedlmair2012taxonomy}
M.~Sedlmair, A.~Tatu, T.~Munzner, and M.~Tory.
\newblock A taxonomy of visual cluster separation factors.
\newblock {\em Computer Graphics Forum}, 31(3pt4):1335--1344, 2012. doi: {{%
10\hspace{.1pt}\discretionary{.}{%
}{.}\hspace{.4pt}1111\discretionary{/}{%
}{/}j\hspace{.1pt}\discretionary{.}{%
}{.}\hspace{.4pt}1467\discretionary{%
}{-}{-}8659\hspace{.1pt}\discretionary{.}{%
}{.}\hspace{.4pt}2012\hspace{.1pt}\discretionary{.}{%
}{.}\hspace{.4pt}03125\hspace{.1pt}\discretionary{.}{%
}{.}\hspace{.4pt}x}}


\bibitem{setlur2016linguistic}
V.~Setlur and M.~C. Stone.
\newblock A linguistic approach to categorical color assignment for data
  visualization.
\newblock {\em IEEE Trans. Vis. \& Comp. Graphics}, 22(1):698--707, 2016. doi:
  {{%
10\hspace{.1pt}\discretionary{.}{%
}{.}\hspace{.4pt}1109\discretionary{/}{%
}{/}tvcg\hspace{.1pt}\discretionary{.}{%
}{.}\hspace{.4pt}2015\hspace{.1pt}\discretionary{.}{%
}{.}\hspace{.4pt}2467471}}


\bibitem{sharma2005ciede2000}
G.~Sharma, W.~Wu, and E.~N. Dalal.
\newblock The ciede2000 color-difference formula: Implementation notes,
  supplementary test data, and mathematical observations.
\newblock {\em Color Research \& Application}, 30(1):21--30, 2005. doi: {{%
10\hspace{.1pt}\discretionary{.}{%
}{.}\hspace{.4pt}1002\discretionary{/}{%
}{/}col\hspace{.1pt}\discretionary{.}{%
}{.}\hspace{.4pt}20070}}


\bibitem{tableau}
{Tableau Software}.
\newblock The tableau visualization system.
\newblock \url{http://www.tableausoftware.com/}.

\bibitem{tominski2008task}
C.~Tominski, G.~Fuchs, and H.~Schumann.
\newblock Task-driven color coding.
\newblock In {\em Proc. Int. Conf. on Information Visualisation}, pp. 373--380,
  2008. doi: {{%
10\hspace{.1pt}\discretionary{.}{%
}{.}\hspace{.4pt}1109\discretionary{/}{%
}{/}iv\hspace{.1pt}\discretionary{.}{%
}{.}\hspace{.4pt}2008\hspace{.1pt}\discretionary{.}{%
}{.}\hspace{.4pt}24}}


\bibitem{tufte1990envisioning}
E.~R. Tufte, N.~H. Goeler, and R.~Benson.
\newblock {\em Envisioning information}, vol. 126.
\newblock Graphics press Cheshire, CT, 1990.

\bibitem{tversky1974judgment}
A.~Tversky and D.~Kahneman.
\newblock Judgment under uncertainty: Heuristics and biases.
\newblock {\em science}, 185(4157):1124--1131, 1974.

\bibitem{veltkamp1992gamma}
R.~C. Veltkamp.
\newblock The $\gamma$-neighborhood graph.
\newblock {\em Computational Geometry}, 1(4):227--246, 1992.

\bibitem{wang2008color}
L.~Wang, J.~Giesen, K.~T. McDonnell, P.~Zolliker, and K.~Mueller.
\newblock Color design for illustrative visualization.
\newblock {\em IEEE Trans. Vis. \& Comp. Graphics}, 14(6):1739--1754, 2008.
  doi: {{%
10\hspace{.1pt}\discretionary{.}{%
}{.}\hspace{.4pt}1109\discretionary{/}{%
}{/}tvcg\hspace{.1pt}\discretionary{.}{%
}{.}\hspace{.4pt}2008\hspace{.1pt}\discretionary{.}{%
}{.}\hspace{.4pt}118}}


\bibitem{wang2019optimizing}
Y.~Wang, X.~Chen, T.~Ge, C.~Bao, M.~Sedlmair, C.-W. Fu, O.~Deussen, and
  B.~Chen.
\newblock Optimizing color assignment for perception of class separability in
  multiclass scatterplots.
\newblock {\em IEEE Trans. Vis. \& Comp. Graphics}, 25(1):820--829, 2019. doi:
  {{%
10\hspace{.1pt}\discretionary{.}{%
}{.}\hspace{.4pt}1109\discretionary{/}{%
}{/}TVCG\hspace{.1pt}\discretionary{.}{%
}{.}\hspace{.4pt}2018\hspace{.1pt}\discretionary{.}{%
}{.}\hspace{.4pt}2864912}}


\bibitem{wijffelaars2008generating}
M.~Wijffelaars, R.~Vliegen, J.~J. Van~Wijk, and E.-J. Van Der~Linden.
\newblock Generating color palettes using intuitive parameters.
\newblock 27(3):743--750, 2008.

\bibitem{zeileis2009escaping}
A.~Zeileis, K.~Hornik, and P.~Murrell.
\newblock Escaping {RGBland}: selecting colors for statistical graphics.
\newblock {\em Computational Statistics \& Data Analysis}, 53(9):3259--3270,
  2009. doi: {{%
10\hspace{.1pt}\discretionary{.}{%
}{.}\hspace{.4pt}1016\discretionary{/}{%
}{/}j\hspace{.1pt}\discretionary{.}{%
}{.}\hspace{.4pt}csda\hspace{.1pt}\discretionary{.}{%
}{.}\hspace{.4pt}2008\hspace{.1pt}\discretionary{.}{%
}{.}\hspace{.4pt}11\hspace{.1pt}\discretionary{.}{%
}{.}\hspace{.4pt}033}}


\bibitem{zhou2016survey}
L.~Zhou and C.~D. Hansen.
\newblock A survey of colormaps in visualization.
\newblock {\em IEEE Trans. Vis. \& Comp. Graphics}, 22(8):2051--2069, 2016.
  doi: {{%
10\hspace{.1pt}\discretionary{.}{%
}{.}\hspace{.4pt}1109\discretionary{/}{%
}{/}tvcg\hspace{.1pt}\discretionary{.}{%
}{.}\hspace{.4pt}2015\hspace{.1pt}\discretionary{.}{%
}{.}\hspace{.4pt}2489649}}


\end{thebibliography}

\end{document}